\documentclass[letterpaper,10pt,twocolumn,final,oneside,conference]{IEEEtran}
\usepackage{multirow}
\usepackage{graphicx}
\usepackage{acronym}
\usepackage{amsmath}
\usepackage{amssymb}
\usepackage{booktabs}
\usepackage{eurosym}
\usepackage{siunitx}
\usepackage[caption=false,font=footnotesize,listofformat=parens]{subfig}
\usepackage{stfloats}
\usepackage{tikz}
\usepackage{hyperref}
\usepackage{multirow}
\usepackage{balance}
\usepackage{comment}
\usepackage{tikz}

\acrodef{dso}[DSO]{distribution system operator}
\acrodef{tso}[TSO]{transmission system operator}
\acrodef{ev}[EV]{electric vehicle}

\DeclareSIUnit{\kWh}{kWh}

\IEEEoverridecommandlockouts

\newcommand\copyrighttext{%
  \footnotesize
  \centering\copyright~2021 IEEE. Personal use of this material is permitted. Permission from IEEE must be obtained for all other uses, in any current or future media, including reprinting/republishing this material for advertising or promotional purposes, creating new collective works, for resale or redistribution to servers or lists, or reuse of any copyrighted component of this work in other works.\\
  IEEE EEEIC/I\&CPS Europe. DOI: \href{https://doi.org/10.1109/EEEIC/ICPSEurope51590.2021.9584782}{10.1109/EEEIC/ICPSEurope51590.2021.9584782}.}
\newcommand\copyrightnotice{%
\begin{tikzpicture}[remember picture,overlay]
\node[anchor=south,yshift=0pt] at (current page.south) {\setlength{\fboxrule}{0pt}\fbox{\parbox{\dimexpr\textwidth-\fboxsep-\fboxrule\relax}{\copyrighttext}}};
\end{tikzpicture}%
}
\begin{document}

\title{Impact of EV Charging Stations in Power Grids in Italy and its Mitigation Mechanisms}

\author{%
  \IEEEauthorblockN{A. Samson Mogos, S. Grillo}
  \IEEEauthorblockA{Politecnico di Milano\\Dipartimento di Elettronica, Informazione e Bioingegneria\\
  p.za Leonardo da Vinci, 32,  I-20133, Milano, Italy\\
    \{aman.samson, samuele.grillo\}@polimi.it}}%

\IEEEaftertitletext{\copyrightnotice\vspace{0.2\baselineskip}}
\maketitle

\begin{abstract}                
Global warming leads the world to think of a different way of transportation: avoiding internal combustion engines and electrifying the transportation sector. With a high penetration of \ac{ev} charging stations on an existing power distribution network, the impact may be consistent. The loads of the fast-charging stations would potentially result in increased peak load demand, reduced reserve margins, voltage instability, and reliability problems. The degrading performance of the power system due to the negative impact of the \ac{ev} charging stations can even lead to penalties to be paid by the \ac{dso}. This paper: i) investigates the impact of the \ac{ev} charging station on the distribution network for what concerns voltage drop on MV feeders and loading of transformers in primary substations, and ii) proposes a mitigation mechanism. A realistic typical Italian grid has been used to assess the impact of \ac{ev} charging stations and to validate the mitigation mechanism.
\end{abstract}


\acresetall
\section{Introduction}

The increasing demand for electricity and the dependence of the energy source on a non-renewable finite fossil fuel supply, accompanied by the impact of pollution on global warming and drastic climate changes, are the key concerns of this era's environmental researchers.

In the transportation sector, nowadays, most of the vehicles are emitting CO\textsubscript{2} which is considered as the main cause for global warming and climate change~\cite{Wang2015,Wang2016}. Most researchers are been engaged in the research of the positive impacts of avoiding internal combustion engine (ICE) driven vehicles and replacing them with \acp{ev} to minimize the carbon emission and protect the climate change and global warming. It is indeed a good idea to shift from conventional vehicles to \acp{ev} as it has many environmental and economic advantages~\cite{Kampman2011}. For this reason, EVs are becoming more popular in the market. However, the increasing number of \acp{ev} is creating a rise in charging load demand. For example, in the USA, the National Program Charging Point America has started to build about 5000 \ac{ev} charging stations in nine regions of the country~\cite{Shi2012}. As per 2020 PHEVs sales data in Europe, it was shown 210\% increase compared to 2019. There was no potential negative effect on the increase of \ac{ev} deployment due to COVID-19~\cite{Falchetta:2021}. It can be easily understood that the world is more aware of global warming and climate change, but not on the impact of \ac{ev} charging stations on the already built power grids.

The installation of \ac{ev} charging stations becomes an additional load burden on the power grids. High \ac{ev} charging loads, especially by fast \ac{ev} charging stations, will decrease the quality of the power system operating parameters on the power distribution grids. The quality of voltage profile will decrease with an increase in the peak load, power system equipment like transformers and power lines will be overloaded during peak hours---which will lead to early aging---, and harmonic distortion is a consequences of the uncoordinated charging of \acp{ev}.

Some researchers have dealt with the impact of \ac{ev} charging loads on different parameters of power system networks like voltage profile~\cite{Geske2010}, harmonics~\cite{Staats1998} and peak load~\cite{McCarthy2010}. One of the potential impacts is the degradation of the voltage profile. In~\cite{Geske2010} authors analyze this issue on a low voltage distribution network in Europe for different \ac{ev} penetration scenarios. They have concluded that the existing network is strong enough to support a low intake of \acp{ev} (i.e., from 1\% to 2\%). It was also concluded that there was a considerable voltage drop on the load nodes where there are many charging stations installed. In~\cite{Juanuwattanakul2011}, authors analyze the impact on a 13-bus distribution network for different \ac{ev} penetration scenarios. They have concluded that the transient voltage stability index degraded for high penetration of \acp{ev}. Similarly, in~\cite{Zhang2016}, the impact of large \ac{ev} charging load penetration on voltage stability on the power distribution network is conducted, concluding that most of the existing power distribution networks can tolerate an \ac{ev} charging load penetration only for a limited level. They both conclude that, having been the networks designed and planned some years ago and assuming that the electric load demand will increase in the coming years, they are inadequate and they are unlikely to withstand high \ac{ev} charging load penetration. Either an enhancement of power system equipment or a re-planning will be required to address this issue.

Researchers in~\cite{Yang2019,Yang2017}, have studied that \ac{ev} charging stations supported by renewable energy have the ability to meet the increasing \acp{ev} power demand and have the potential to mitigate their impact in power grids. Authors in~\cite{Verma2020} have designed a control strategy to utilize different energy sources to minimize the  grid congestion due to \acp{ev} charging load.\IEEEpubidadjcol

Most of the previously conducted studies are focused mainly on impacts of \ac{ev} charging station loads such as voltage instability, harmonic distortion, and power losses on the distribution network. Furthermore, most of the studies were conducted on test networks. Very few scholars have performed their research on a real built-in power distribution network. Considering the aforementioned statements this paper is conducted on a realistic and typical power distribution network of the industrial belt of a large Italian city. Some of the major contributions of the study are the following:
\begin{itemize}
  \item A brief analysis of the impact of the \ac{ev} charging station loads on the voltage stability (maximum voltage variation) of the distribution network.
  \item Detailed analysis of the impact of the \ac{ev} charging station loads on the power system elements loading and peak load.
  \item Forecasted analysis of the \ac{ev} charging load on different parameters of the distribution network such as the voltage stability, peak load, and equipment loading, at the years of 2020, 2023, 2025, 2026 and 2030.
  \item Applying the mitigation mechanism and evaluating the grid characteristics.
\end{itemize}

The paper is organized as follows. In Section II the grid under study is described. In Section III the different case studies used for analyzing the impact of EV charge load and testing the effectiveness of the proposed mitigation mechanisms. Section IV is devoted to the simulation of the results. Finally, conclusions are drawn in Section V.

\section{Description of the Grid Under Study}

The grid under study mimics a realistic and typical distribution network in Italy. Moreover, it could be also representative of a more general European distribution system. Primary substations, especially when serving an industrial zone, have multiple busbars feeding several MV portions of the distribution system. In the present study, only one busbar from the main substation is considered. It is composed of 7 MV lines, one \SI{40}{\mega\volt\ampere} HV/LV transformer, 38 MV customers with total installed power of \SI{27.7}{\mega\watt}, distributed generators summing up to \SI{5.4}{\mega\watt} nominal active power and connected to the MV lines, approximately 10,000 LV customers, and 138 secondary substations.

The nominal voltage of the HV system is \SI{132}{\kilo\volt} and the distribution voltage level are \SI{15}{\kilo\volt}. Power transformers are with different power rating ranging from \SI{63}{\kilo\volt\ampere} to \SI{630}{\kilo\volt\ampere}. All the transformers are step-down transformers from \SI{15}{\kilo\volt} to \SI{0.4}{\kilo\volt}, with the obvious exception of the substation transformer (\SI{132/15.6}{\kilo\volt}) and of the step-up generator transformers (\SI{0.4/15}{\kilo\volt}). A piece of summarized quantitative information could be found in the Tables~\ref{tab:distr}--\ref{tab:ders}.

\begin{table}[h]
  \centering
  \caption{Details of the distribution transformer.}
  \label{tab:distr}
  \begin{tabular}{cccc}
  \toprule
  Rating & Cooling & $V_1/V_2$ & Quantity\\
  $[\si{\kilo\volt\ampere}]$ & &  $[\si{\kilo\volt} / \si{\kilo\volt}]$ &\\
  \midrule
  4000&\multirow{7}{*}{ONAN}&132/15.6&1\\
  630&&\multirow{6}{*}{15/0.4}&7\\
  400&&&56\\
  250&&&52\\
  160&&&12\\
  100&&&9\\
  63&&&2\\
  \bottomrule
  \end{tabular}
\end{table}

\begin{table}[h]
  \centering
  \caption{Details of the generator transformer.}
  \label{tab:gentr}
  \begin{tabular}{cccc}
  \toprule
  Rating & Cooling & $V_1/V_2$ & Quantity\\
  $[\si{\kilo\volt\ampere}]$ & &  $[\si{\kilo\volt} / \si{\kilo\volt}]$ &\\
  \midrule
  2000	& \multirow{9}{*}{ONAN} & \multirow{9}{*}{0.4/15} & 1\\
  1600		&&&4\\
  1250		&&&1\\
  1000		&&&3\\
  800		&&&2\\
  630		&&&4\\
  600		&&&1\\
  500		&&&3\\
  400		&&&2\\

  \bottomrule
  \end{tabular}
\end{table}

The distribution branches are overhead or underground cables. In total, there are 178 branch lines with a cross-sectional area ranging from \SI{20}{\milli\metre\squared} to \SI{240}{\milli\metre\squared}. The types of branches are aluminum (AL), aluminum alloy (AC), and copper (CU).

\begin{table}[h]
  \centering
  \caption{Details of the branches.}
  \label{tab:branches}
  \begin{tabular}{cccc}
  \toprule
  Conductor size & Material & Total length & Quantity\\
  $[\si{\milli\metre\squared}]$ & &  $[\si{\kilo\metre}]$ &\\
  \midrule
  240 & AL & 18.7 & 49\\
  185 & AL & 16.8 & 39\\
  150 & AL/CU & 20.1 & 47\\
  100 & CU & 0.337 & 2\\
  95 & CU & 4.67 & 19\\
  70 & CU & 0.16 & 1\\
  63 & CU & 0.44 & 2\\
  54 & AC & 3.34 & 2\\
  40 & CU & 0.72 & 4\\
  35 & AL & 2.16 & 6\\
  25 & CU & 1.15 & 5\\
  20 & CU & 0.07 & 1\\
  \bottomrule
  \end{tabular}
\end{table}

Apart from the main generation of the national grid, there are some static and dynamic generation plants connected to the grid under study. The generators are Synchronous (SI), Asynchronous (AS), and static (ST) type. All types of static generators are solar PV power plants. The nominal voltage of all the generators is \SI{0.4}{\kilo\volt}. The power factor of all the solar PV generators is considered to be 1, while it is either 0.8 or 0.9 for the dynamic generators. A summarized quantitative data of the generators could be found in Table IV.

\begin{table}[h]
  \centering
  \caption{Details of the distributed generators.}
  \label{tab:ders}
  \begin{tabular}{cccc}
  \toprule
  Gen. Type & Plant Type & Rated Power & Quantity\\
   & &  $[\si{\kilo\volt\ampere}]$ &\\
  \midrule
  SI	& Genset & 2842 & 3\\
  AS	& Induction & 829 & 7\\
  ST	& PV & 4685.8 & 11\\
  \bottomrule
  \end{tabular}
\end{table}

\begin{figure}
  \centering
  \begin{tabular}{cc}
  \subfloat[][\label{fig:fig_1_1}]{\includegraphics[width=.46\columnwidth]{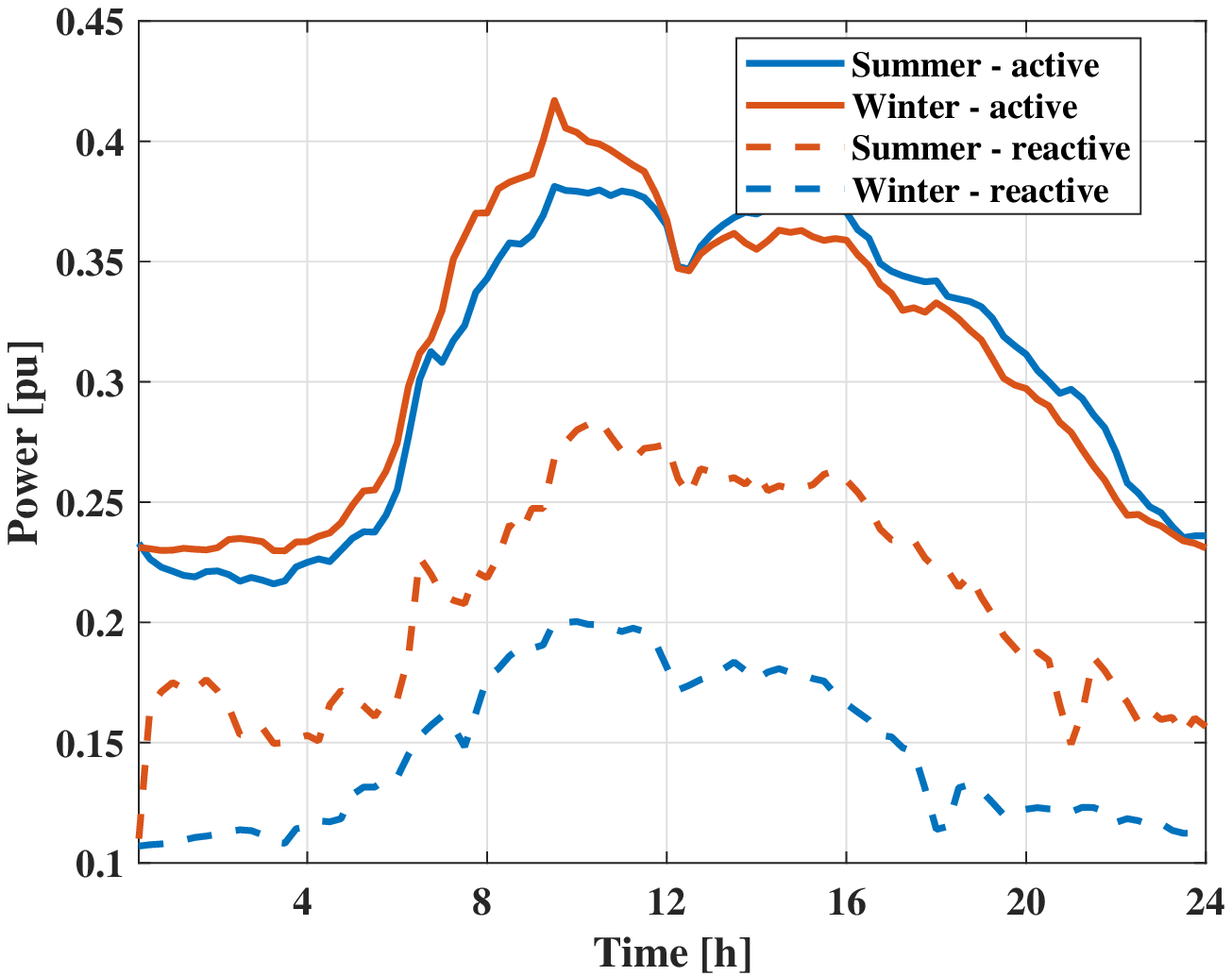}}&
  \subfloat[][\label{fig:fig_1_2}]{\includegraphics[width=.46\columnwidth]{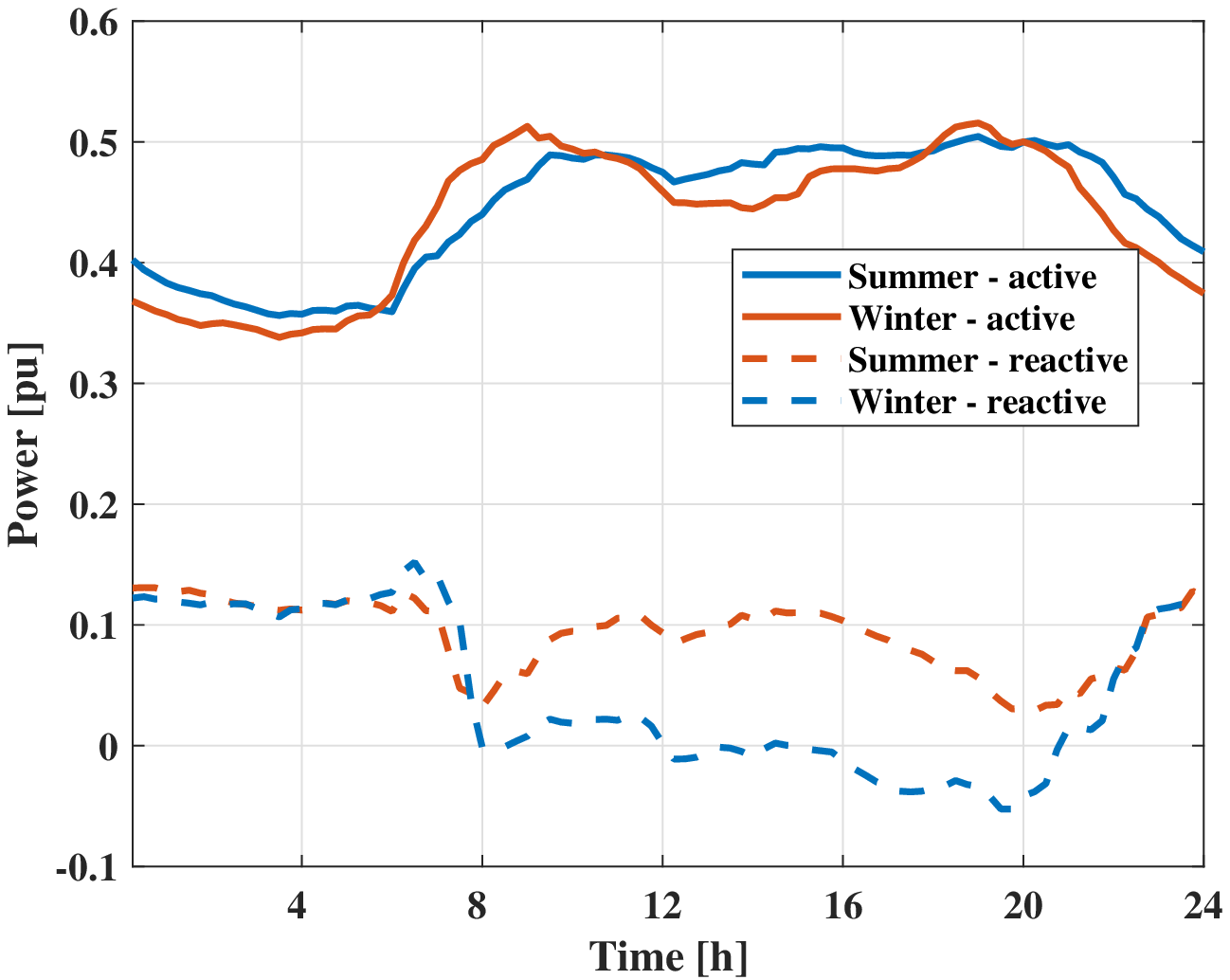}}
  \end{tabular}
  \caption{Load profiles used in the present study: (a) MV customers, and (b) LV customers.}\label{fig:load_1}
\end{figure}

The MV clients that are connected to this grid are mainly industries. The total installed power of the MV customers is \SI{27.7}{\mega\volt\ampere}. Some of the MV clients are not only consumers but also generators. Thirteen out of the 38 MV customers sum up to a total of \SI{13.3}{\mega\watt} installed nominal power and they produce up to \SI{5.4}{\mega\watt} in total. In this study, the average load profile of the MV customers is considered. The load profile was calculated based on a real-time measurements provided by the DSO of the months of February, May, July, and November. The load profile was taken every \SI{15}{\minute} in a day for a full month. Figure~\subref{fig:fig_1_1} shows the average MV per unit load profiles. The load measured in February has been taken as ``winter load'', while the one in July as ``summer load''.

There are more than 10,000 LV customer connected to the network by means of 153 MV/LV transformers in the secondary substations. All LV customers under the same secondary substation have been gathered in one single equivalent three-phase load. For the ease of the study, the LV and MV loads are considered as a balanced load. The average LV load profile considered are reported in Fig.~\subref{fig:fig_1_2}.

\section{Case Studies}
The following case studies are conducted to investigate the impacts of \ac{ev} load on the power grid. Two types of \ac{ev} loads are considered, the ones connected to chargers on households and the ones connected to a park \& ride (P\&R) facility. In this study two cars per household (LV customer) is considered~\cite{Shafiee2013}. Although not all \ac{ev} charging stations have power factor corrector, most of them operate at a unity power factor unless they are used for reactive power compensation. In this paper all \ac{ev} charging loads are considered to have a unity power factor.

In the following four case studies, the effect of different \ac{ev} load penetration levels on the performance of power distribution grid, i.e., voltage profile deviation, percentage of loading of the main transformer, and peak to average ratio (PAR), are assessed. The \ac{ev} load penetration considered on this paper are according~\cite{Shafiee2013}.

\subsection{Case I}

This is a single year study for the year of 2020. Penetration levels of 11\% (low), 35\% (medium), and 45\% (high) are considered~\cite{Shafiee2013}. In this case only the \ac{ev} loads of households are considered.

\subsection{Case II}

In this case, three definite years are selected to probe the impacts of medium \ac{ev} penetration level for the year of 2020, 2023 and 2026.
\begin{itemize}
  \item The year 2020: \ac{ev} penetration level is 35\% and the LV load profiles reported on Figures 3 and 4 are considered for each household.
  \item The year 2023: \ac{ev} penetration level is 47\% and other load profiles show a load growth of 4.9\% compared to that of 2020.
  \item The year 2026: \ac{ev} penetration level is 50\% and other load profiles show a load growth of 7.8\% compared to that of 2020.
\end{itemize}

The assumption made here is that the infrastructure of power grid remains the same, while the number of LV and MV customers increases by 1.3\% each year due to the growth of population ~\cite{Conti2007,Shafiee2013} and other factors like industrial growth.

\subsection{Case III}

This case is a study conducted for 2020 with a medium household \ac{ev} load penetration and an additional \ac{ev} charger on a P\&R facility. Three P\&R facilities with 1000 parking lots each equipped with \SI{3.3}{\kilo\watt} \ac{ev} charger is considered. This makes the total installed \ac{ev} load in the P\&R facilities to \SI{9.9}{\mega\watt}.

\subsection{Case IV}

This case could be the most likely to happen in the future with all \ac{ev} loads considered. The load growth will be considered 0.9\% per year instead of 1.3\%. The 0.9\% load growth is the most likely values in Italy~\cite{EPRI:2007}. Some probable \ac{ev} penetration percentages and years are taken. This case is aimed to study the long-term impact: 10\% \ac{ev} load penetration in 2020, 30\% \ac{ev} load penetration in 2025 and 50\% \ac{ev} load penetration in 2030.

\section{Simulation and Results}

Load flow analysis was performed to calculate the total load and voltage deviation in different load points of the network using DIgSILENT Power Factory. Power flow is simulated sequentially over successive time steps using the quasi-dynamic simulation function. This Section presents the results obtained when the simulations are run for the case studies reported in Section~III with unoptimized and optimized \ac{ev} charger loads. It also discusses the result obtained in reducing the negative impact of \ac{ev} charger loads when upgrading the main transformer.

The formulation of the \ac{ev} charging load and their optimization mechanism are described in~\cite{Brenna2016}.

\begin{figure}[ht]
  \centering
  \includegraphics[width=.7\columnwidth]{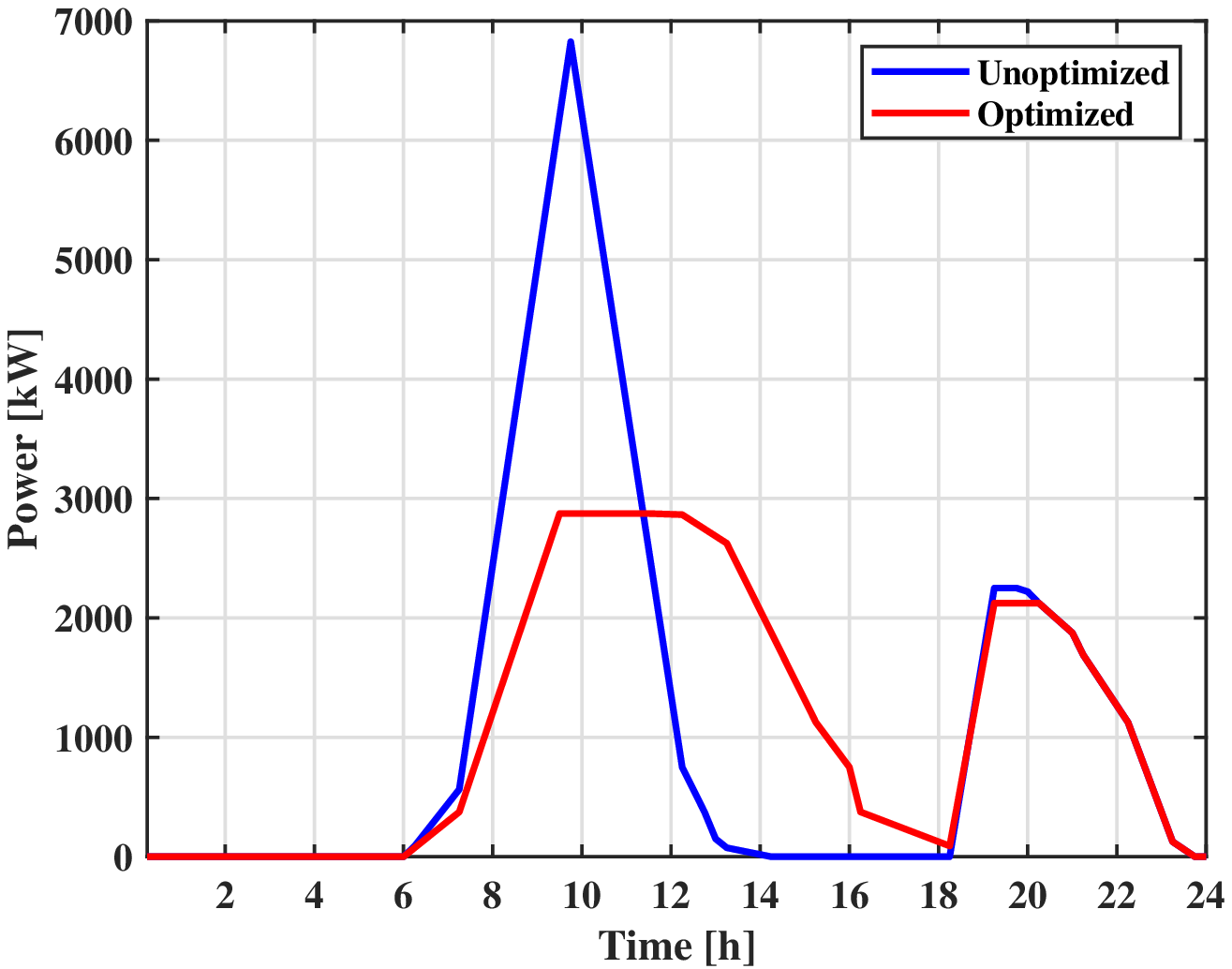}
  \caption{Optimized and unoptimized EV charger load of P\&R facility.}\label{fig:fig_7}
\end{figure}

The objective of the optimization is to charge \acp{ev}, keeping, during each time interval, the power drawn by the $i$-th \ac{ev} below the maximum charging limit of a charge point and the total power of the P\&R facility below the nominal power. The details of the optimization algorithm can be found in~\cite{Brenna2016}. The \ac{ev} load of the P\&R facility is represented in Fig.~\ref{fig:fig_7}, with and without considering the optimization algorithm.

\begin{figure*}[p]
  \centering
  \renewcommand{\arraystretch}{1.3}
  \begin{tabular}{cccc}
    \multicolumn{4}{c}{Case I}\\
    \subfloat[][Voltage profile (summer).\label{fig:fig_8_1}]{\includegraphics[width=.46\columnwidth]{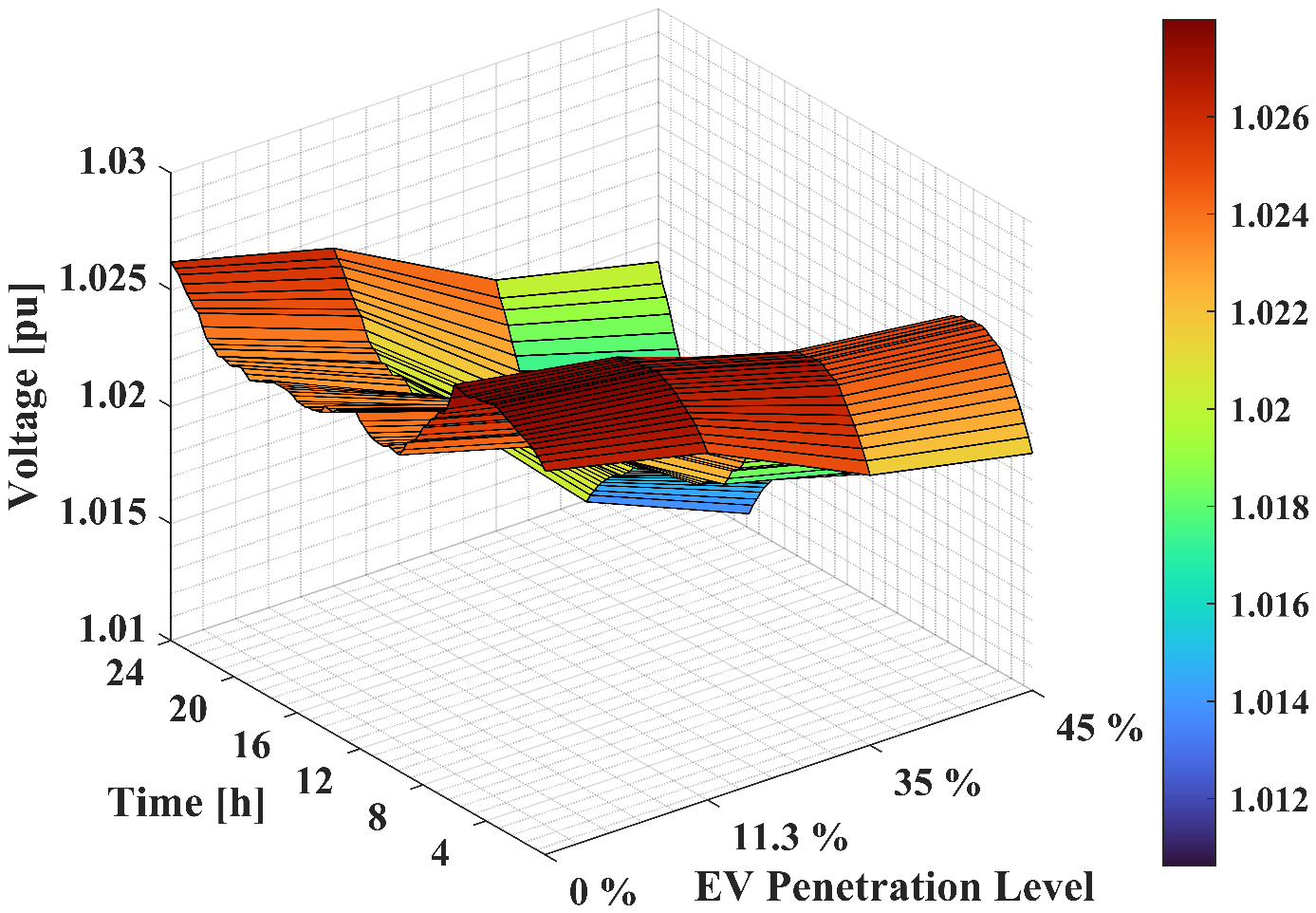}} &
    \subfloat[][Voltage profile (winter).\label{fig:fig_8_2}]{\includegraphics[width=.46\columnwidth]{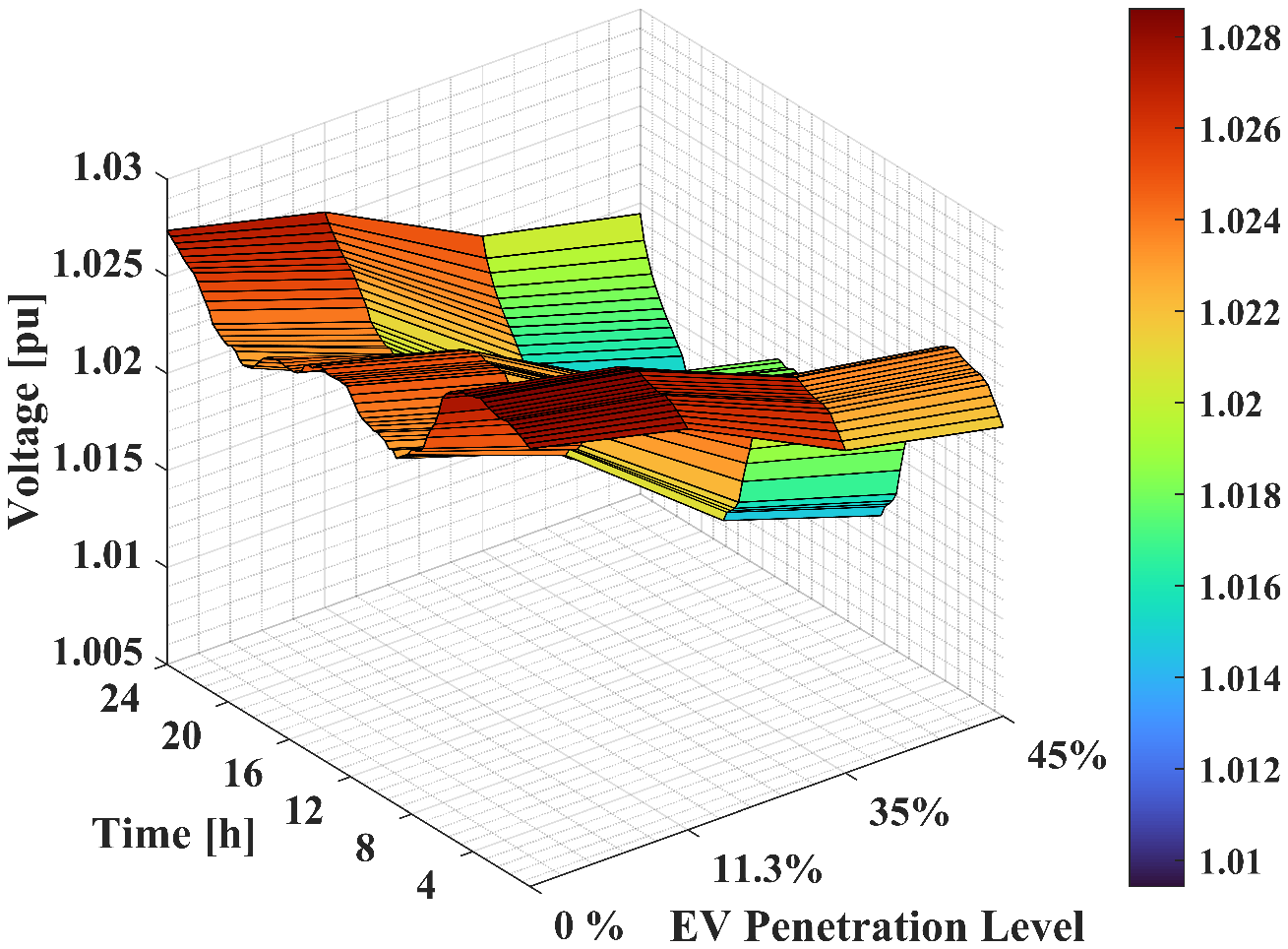}}
    &
    \subfloat[][Trafo loading (summer).\label{fig:fig_9_1}]{\includegraphics[width=.46\columnwidth]{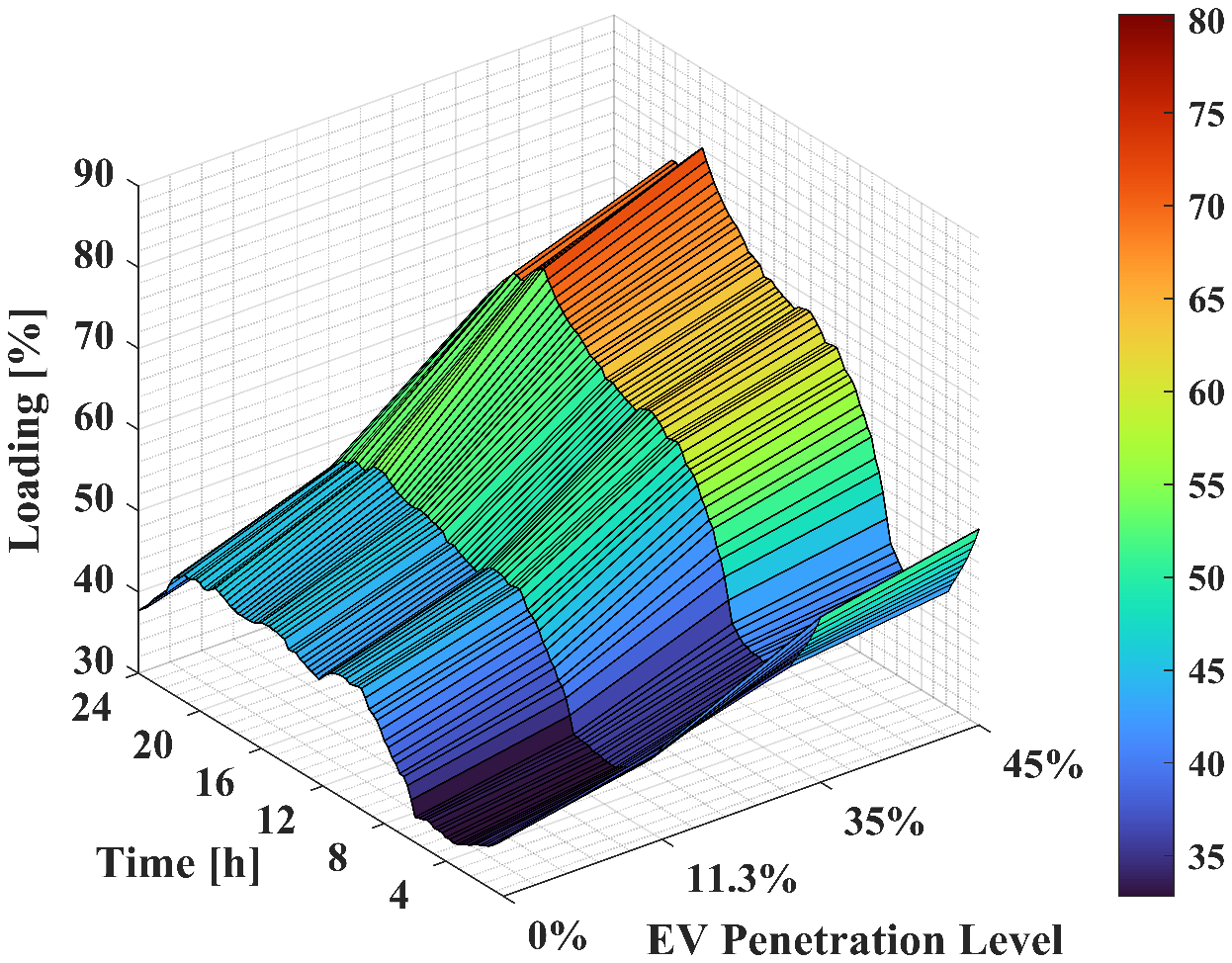}}
    &
    \subfloat[][Trafo loading (winter).\label{fig:fig_9_2}]{\includegraphics[width=.46\columnwidth]{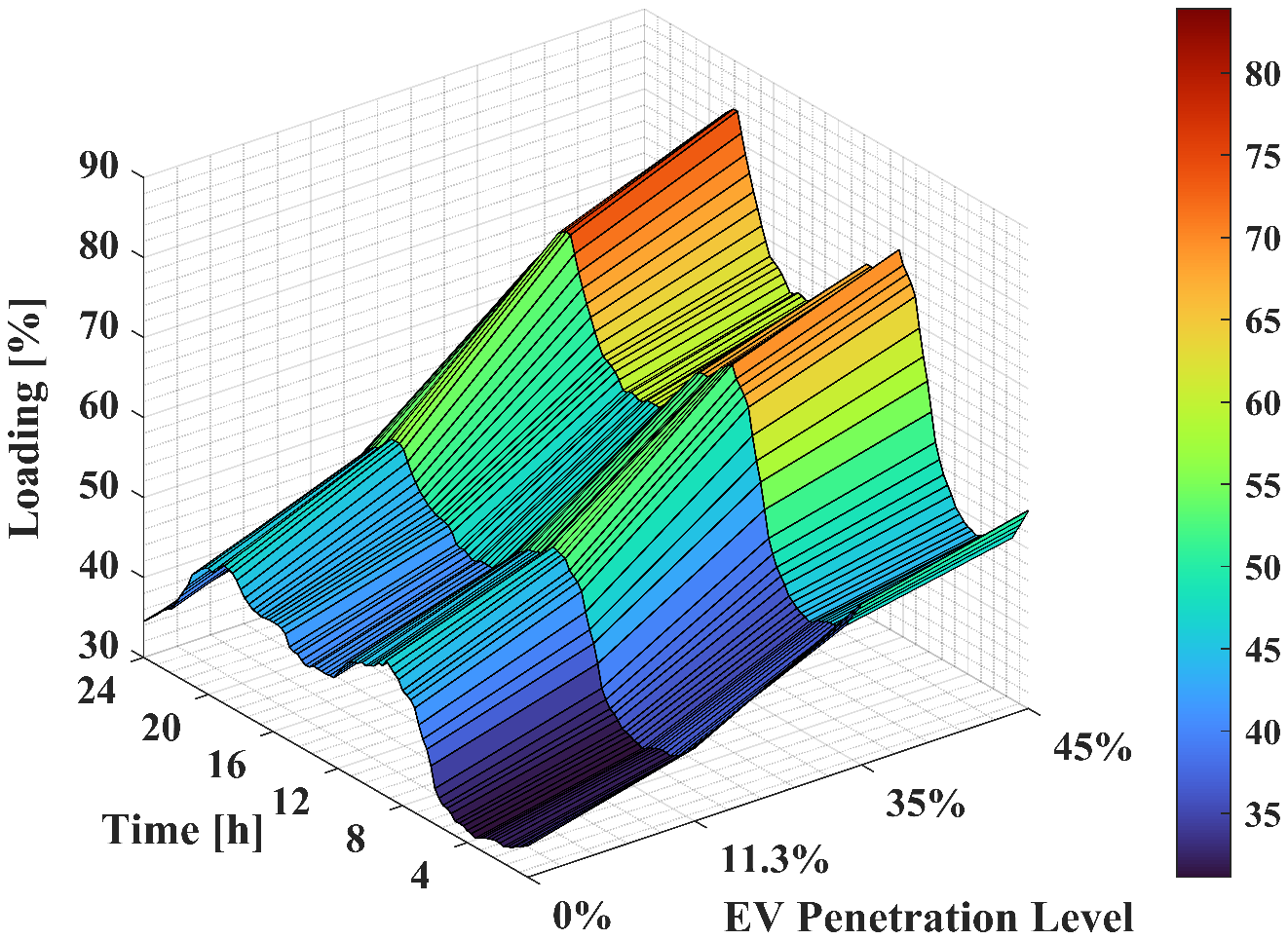}}
    \\
    \multicolumn{4}{c}{Case II}\\
    \subfloat[][Voltage profile (summer).\label{fig:fig_10_1}]{\includegraphics[width=.46\columnwidth]{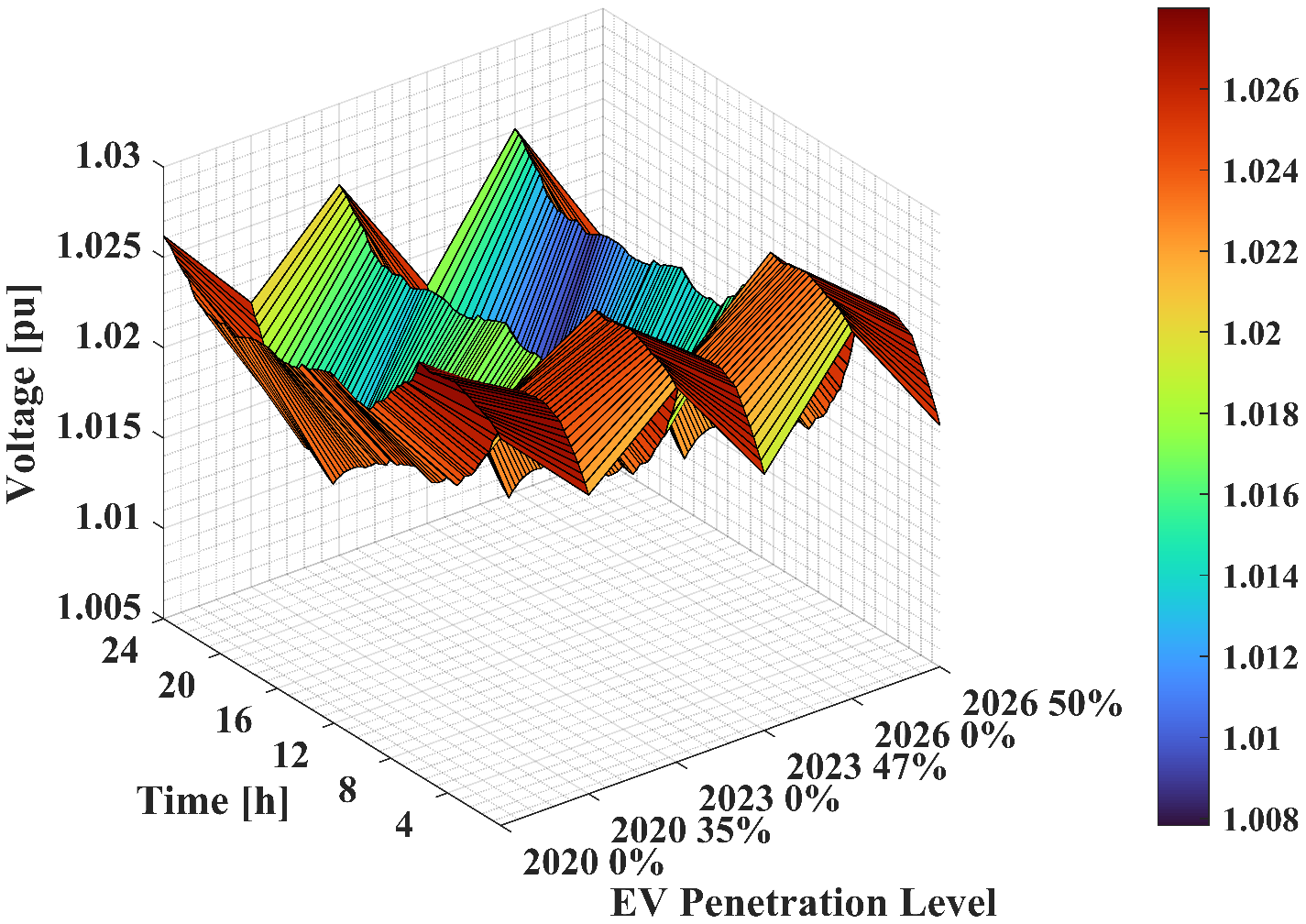}} &
    \subfloat[][Voltage profile (winter).\label{fig:fig_10_2}]{\includegraphics[width=.46\columnwidth]{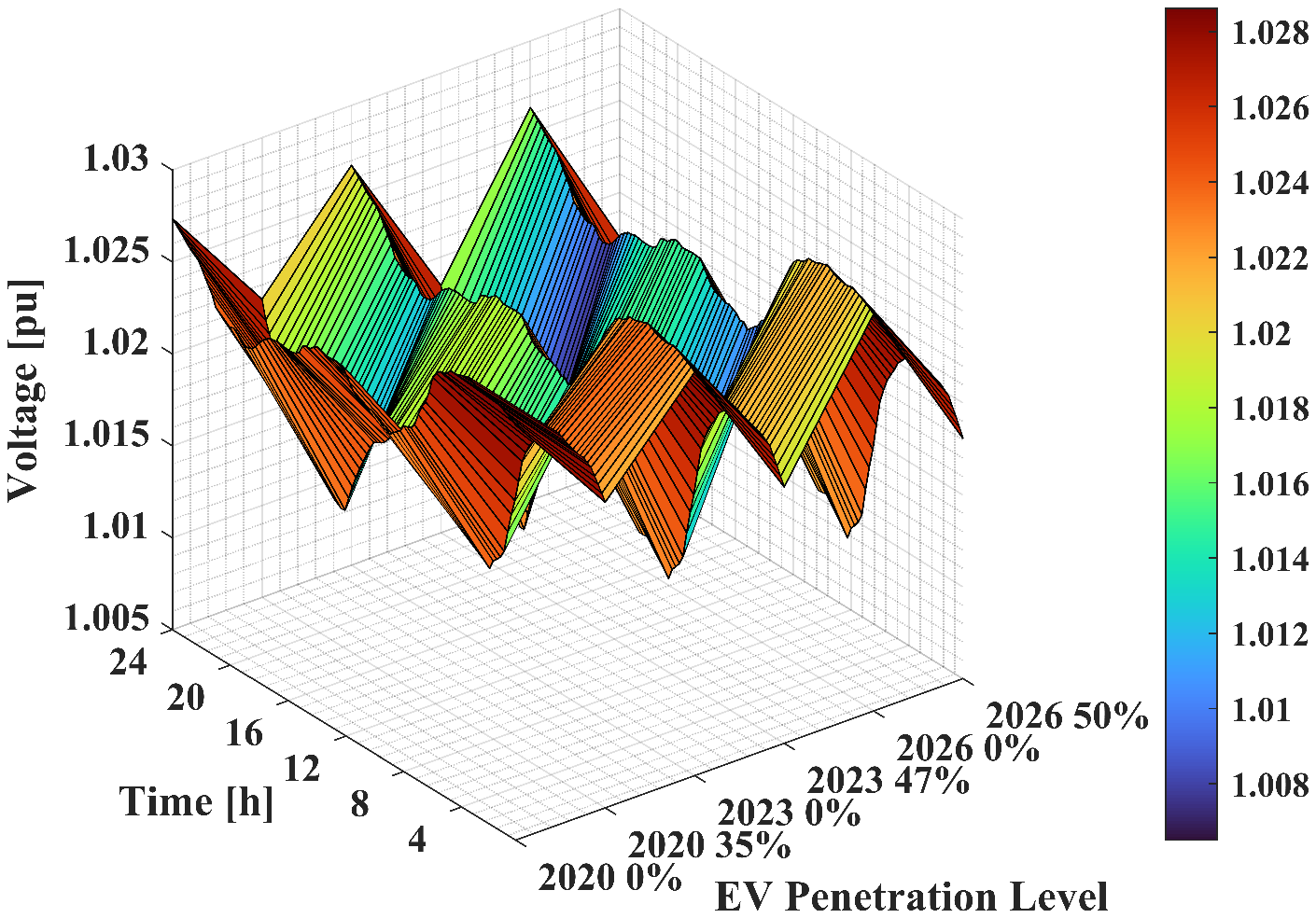}}
    &
    \subfloat[][Trafo loading (summer).\label{fig:fig_11_1}]{\includegraphics[width=.46\columnwidth]{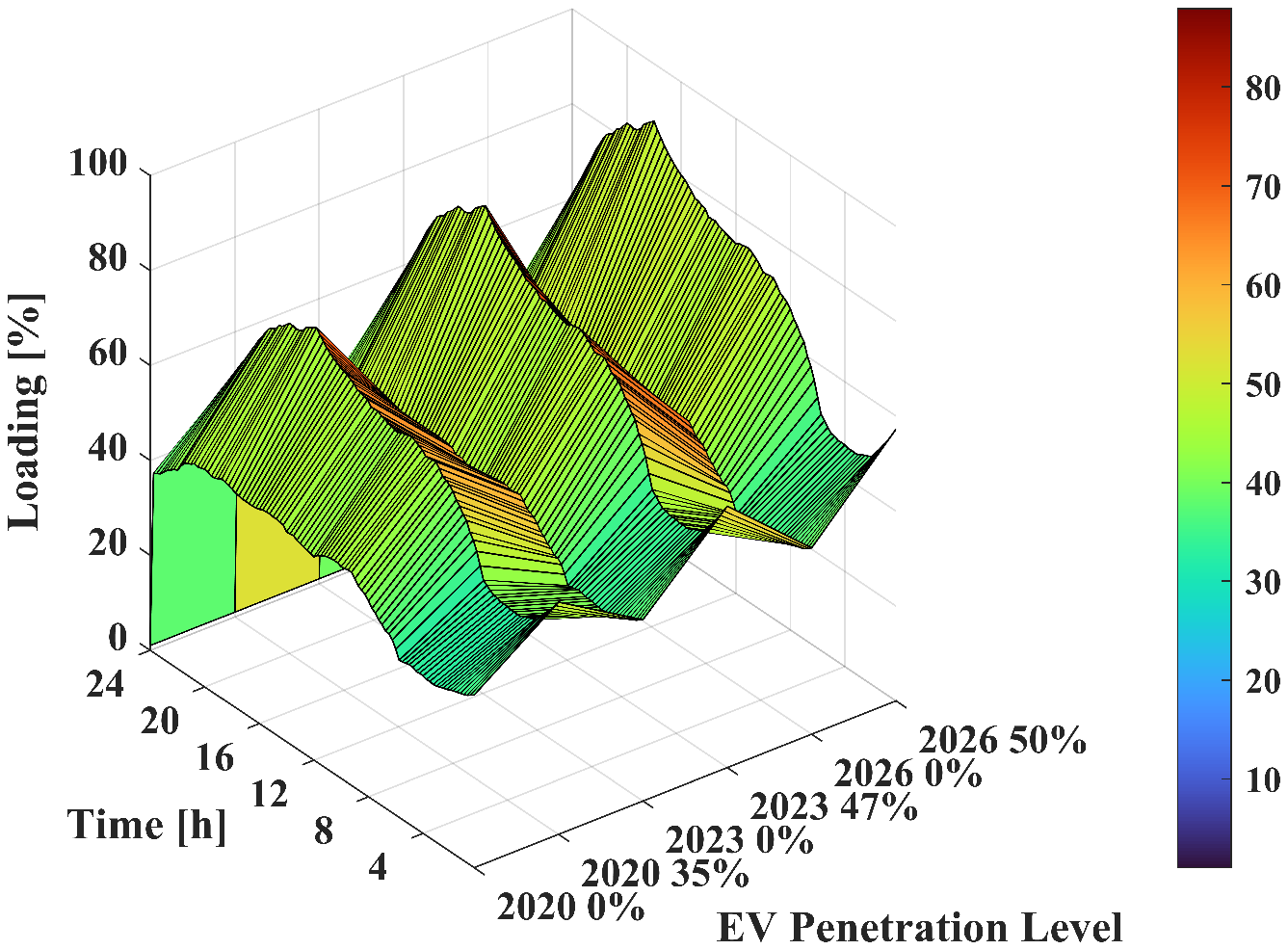}}
    &
    \subfloat[][Trafo loading (winter).\label{fig:fig_11_2}]{\includegraphics[width=.46\columnwidth]{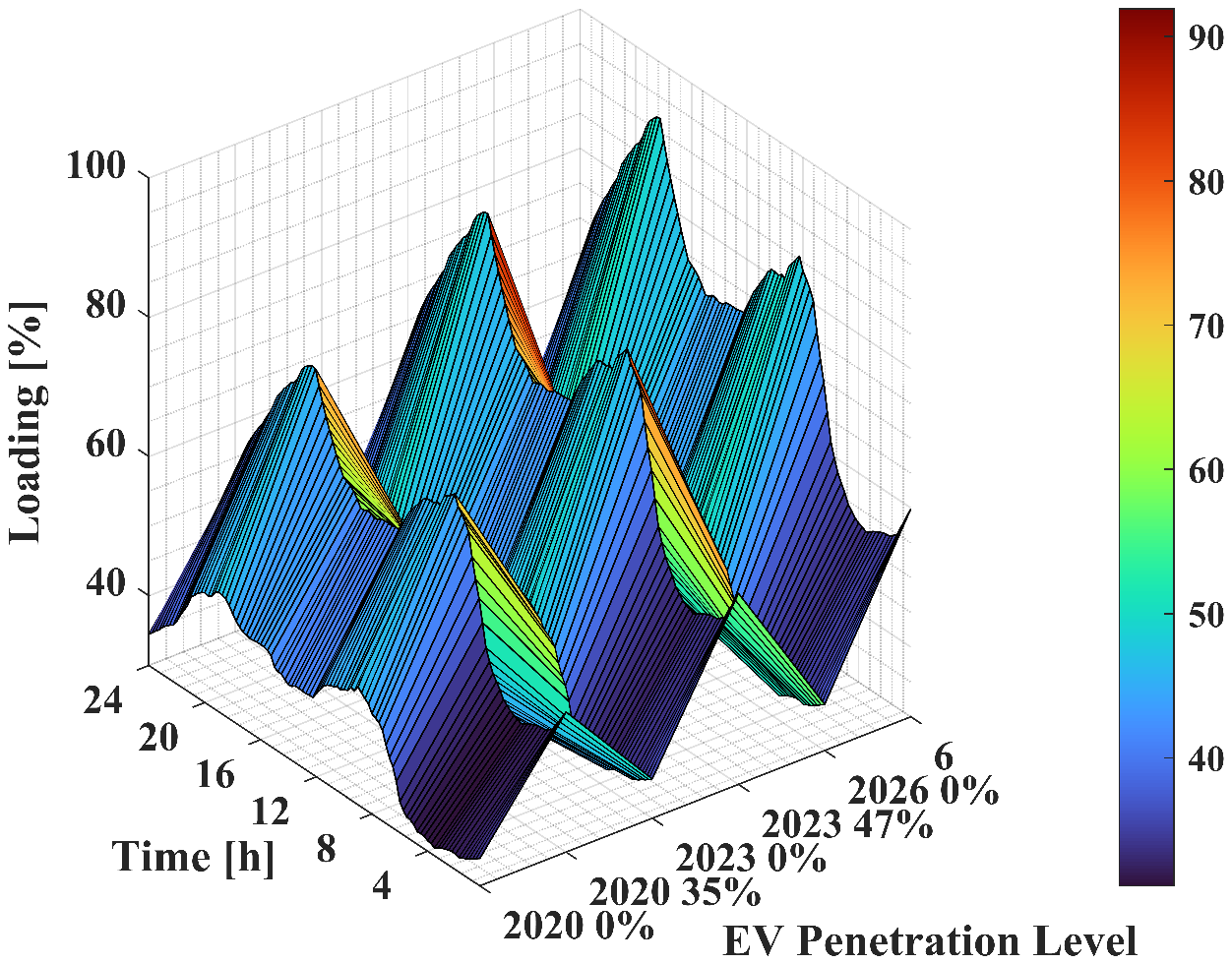}}\\
    \multicolumn{4}{c}{Case III}\\
    \subfloat[][Voltage profile (summer).\label{fig:fig_12_1}]{\includegraphics[width=.46\columnwidth]{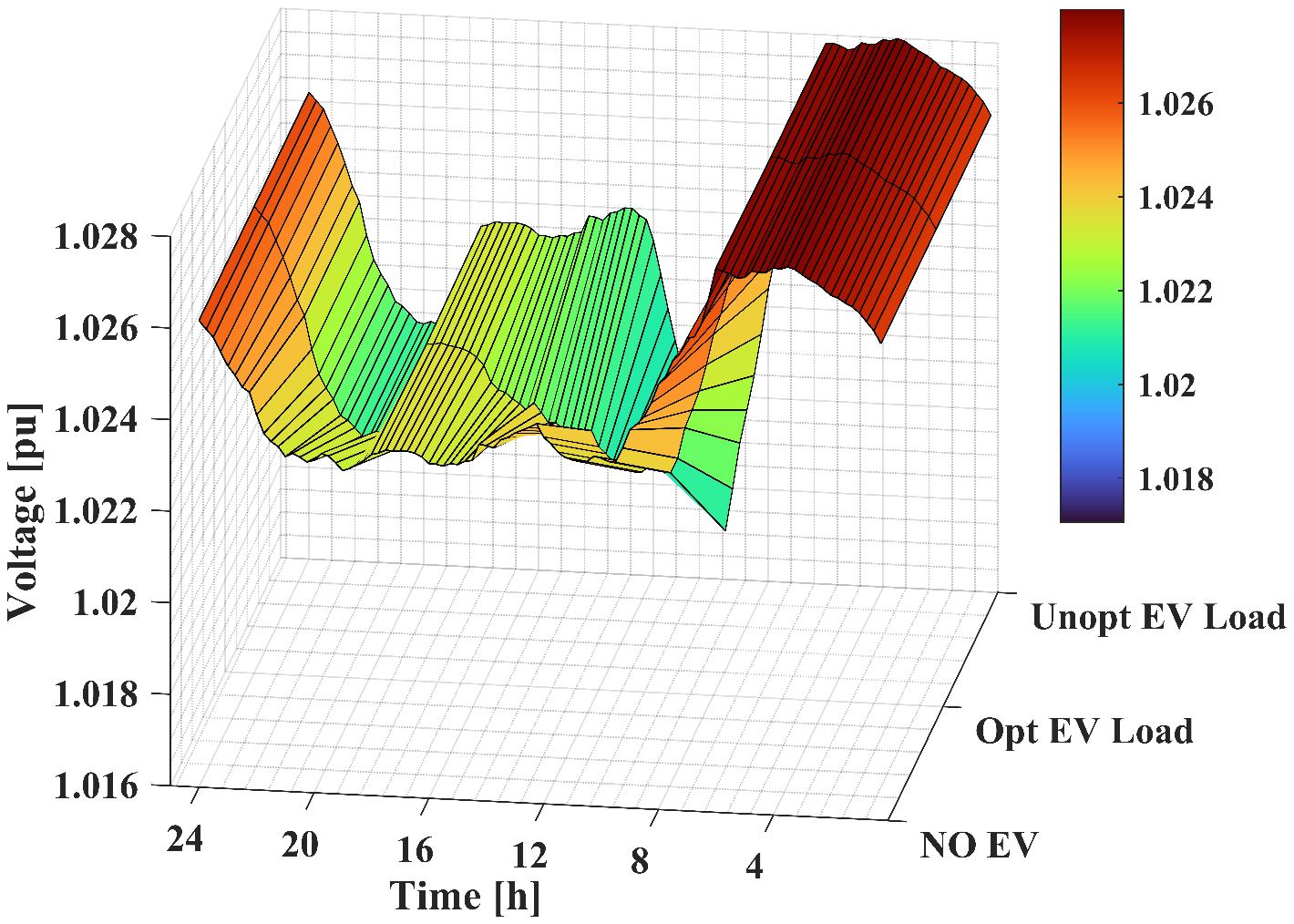}} &
    \subfloat[][Voltage profile (winter).\label{fig:fig_12_2}]{\includegraphics[width=.46\columnwidth]{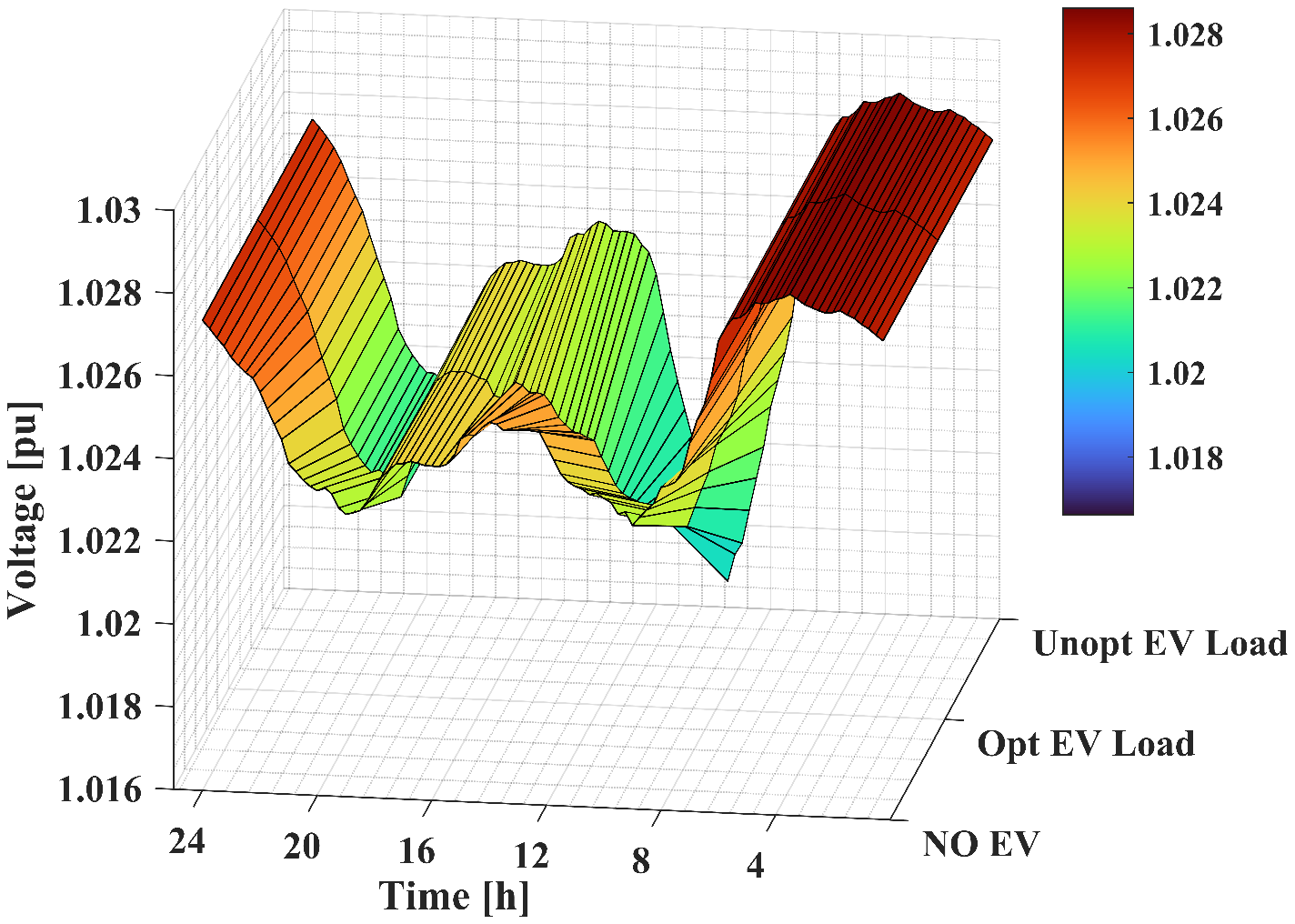}}
    &
    \subfloat[][Trafo loading (summer).\label{fig:fig_13_1}]{\includegraphics[width=.46\columnwidth]{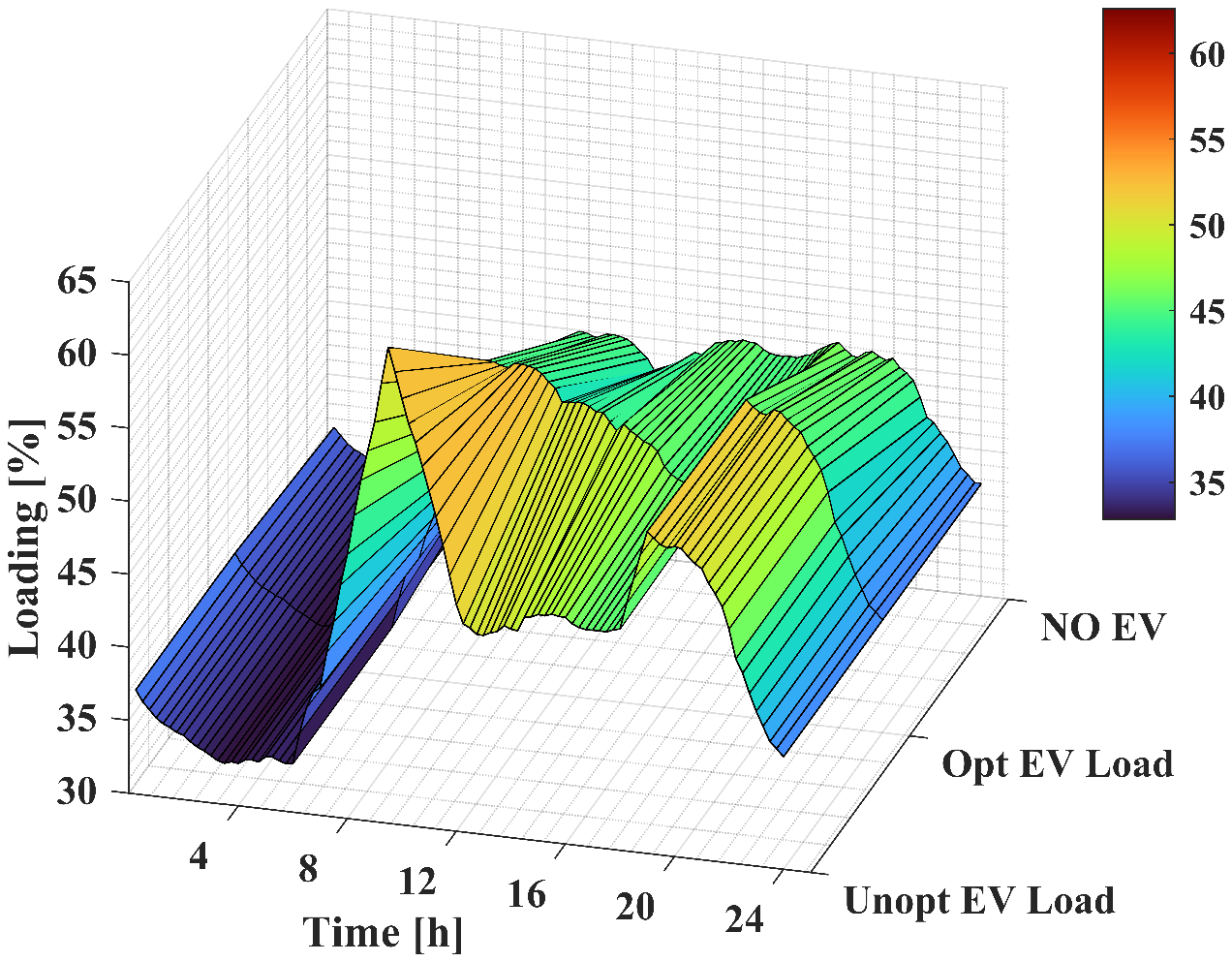}}
    &
    \subfloat[][Trafo loading (winter).\label{fig:fig_13_2}]{\includegraphics[width=.46\columnwidth]{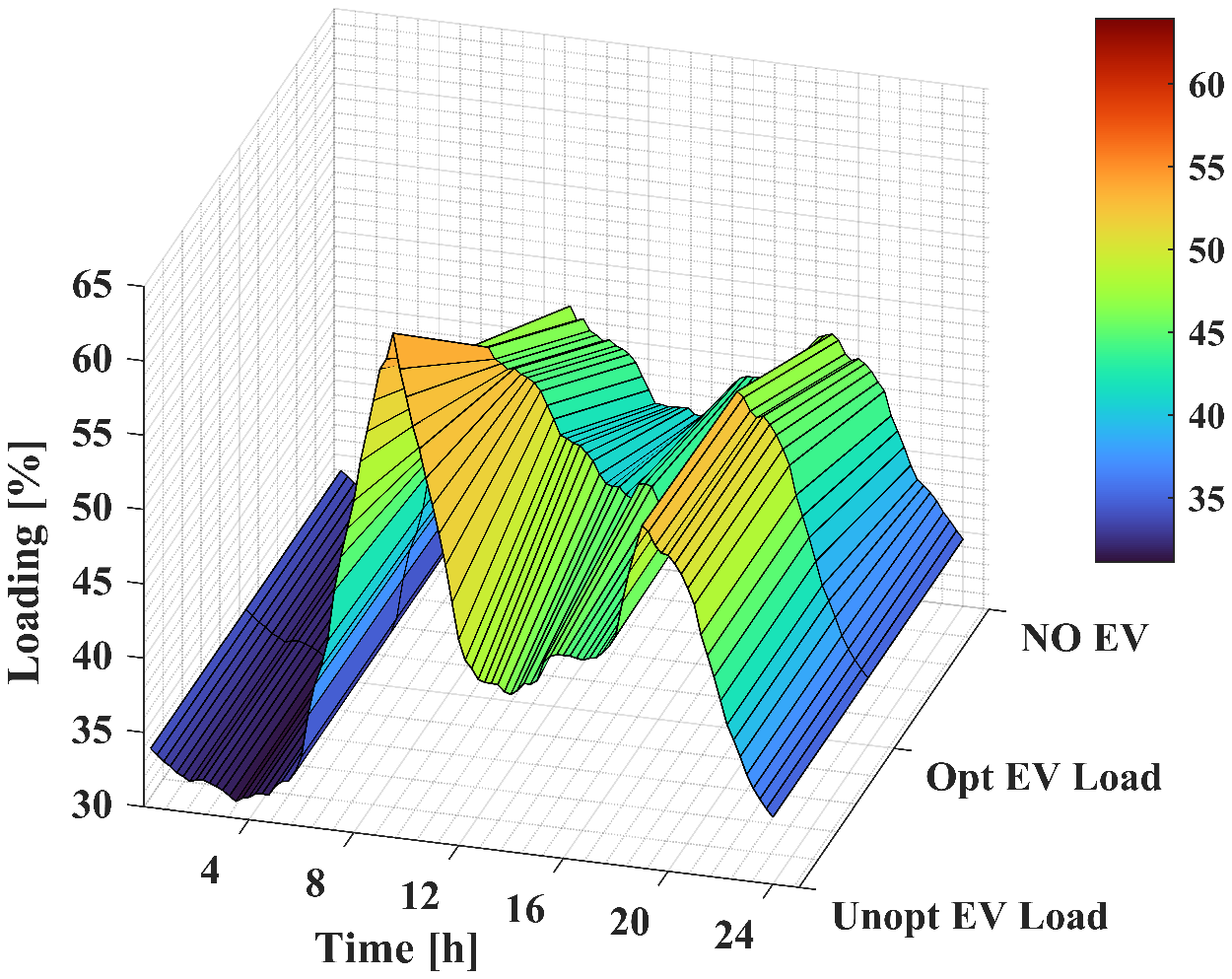}}\\
    \multicolumn{4}{c}{Case IV}\\
    \subfloat[][Voltage profile (summer optimized).\label{fig:fig_14_1}]{\includegraphics[width=.46\columnwidth]{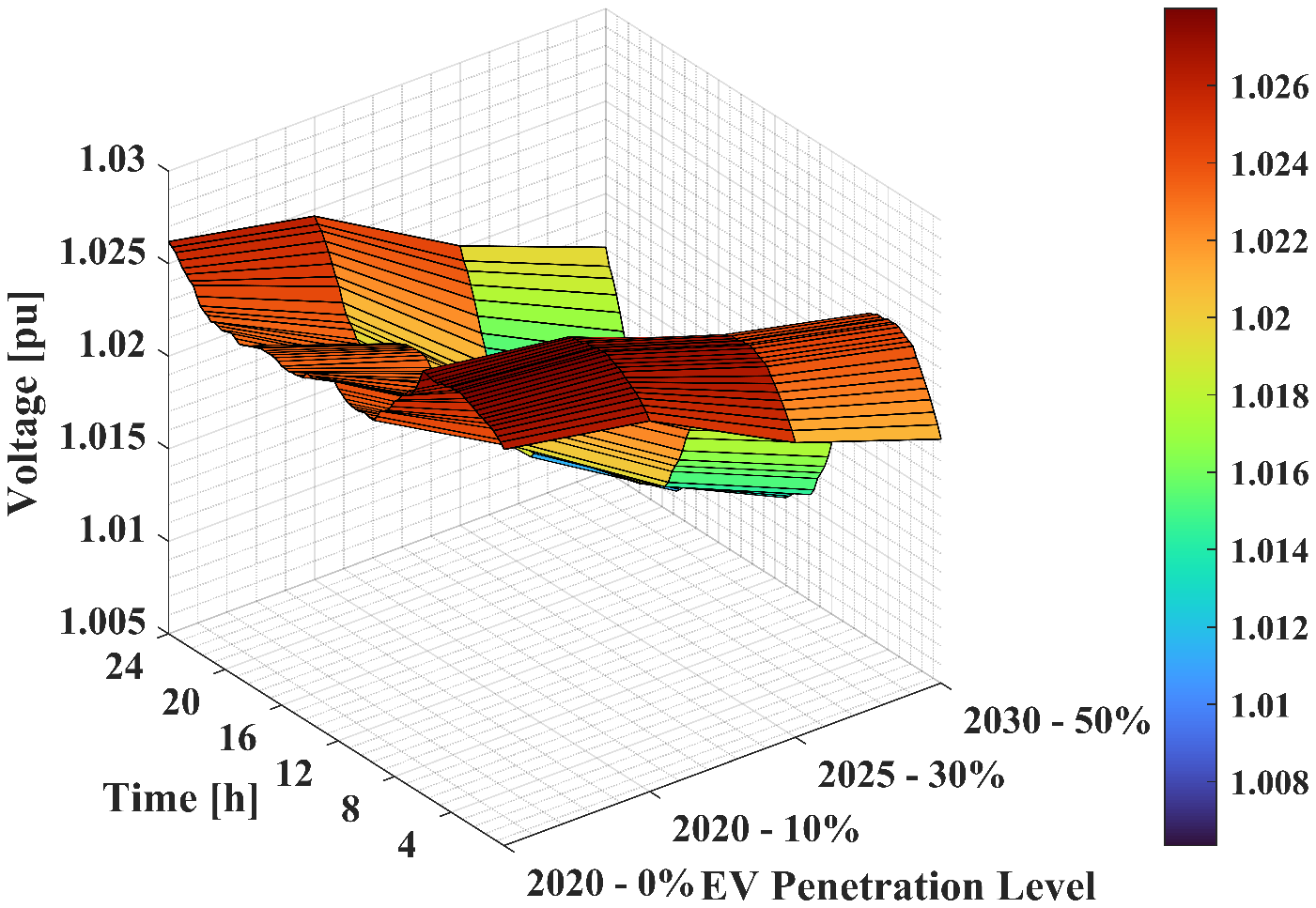}} &
    \subfloat[][Voltage profile (summer unoptimized).\label{fig:fig_14_2}]{\includegraphics[width=.46\columnwidth]{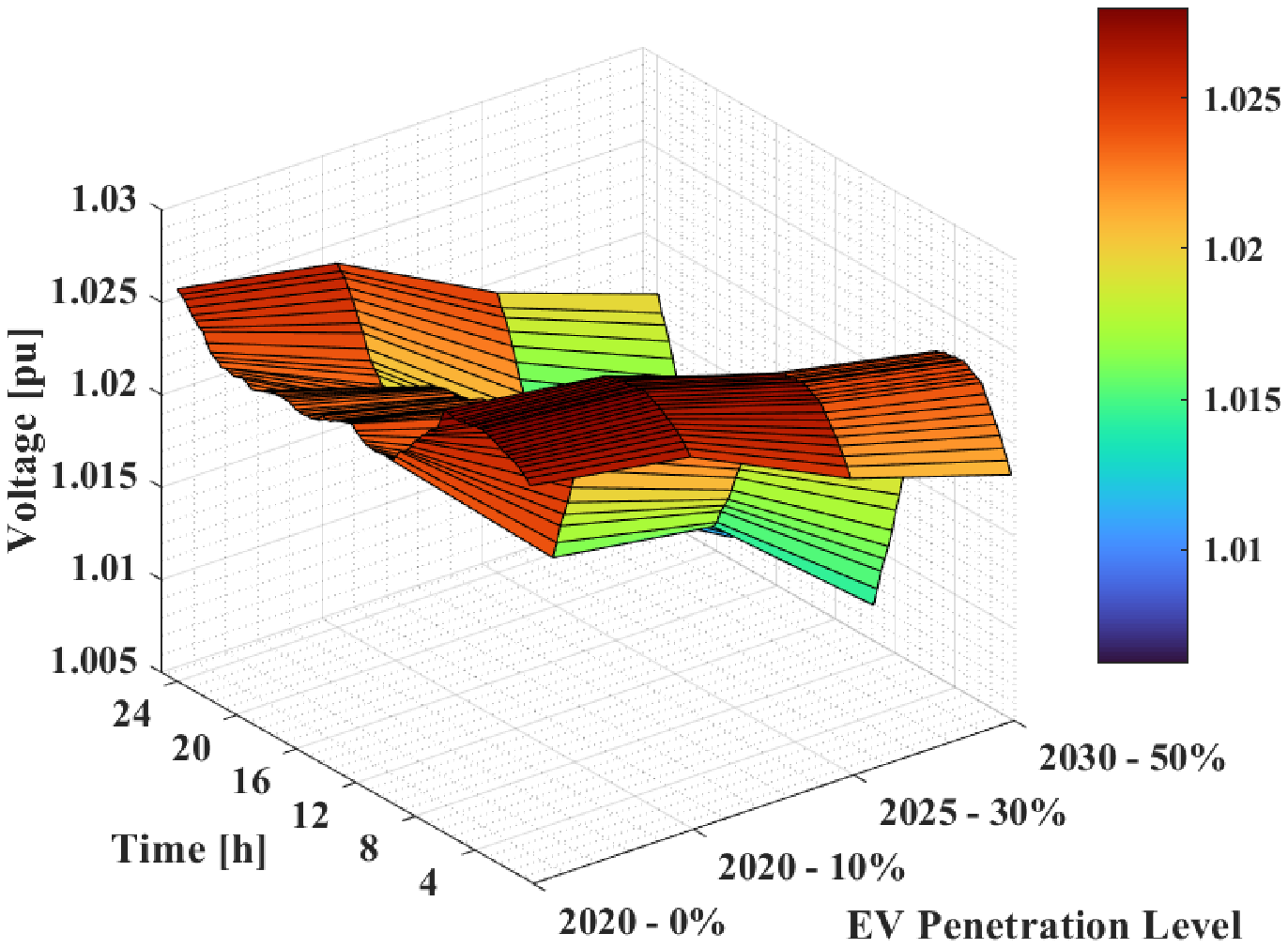}}
    &
    \subfloat[][Voltage profile (winter optimized).\label{fig:fig_15_1}]{\includegraphics[width=.46\columnwidth]{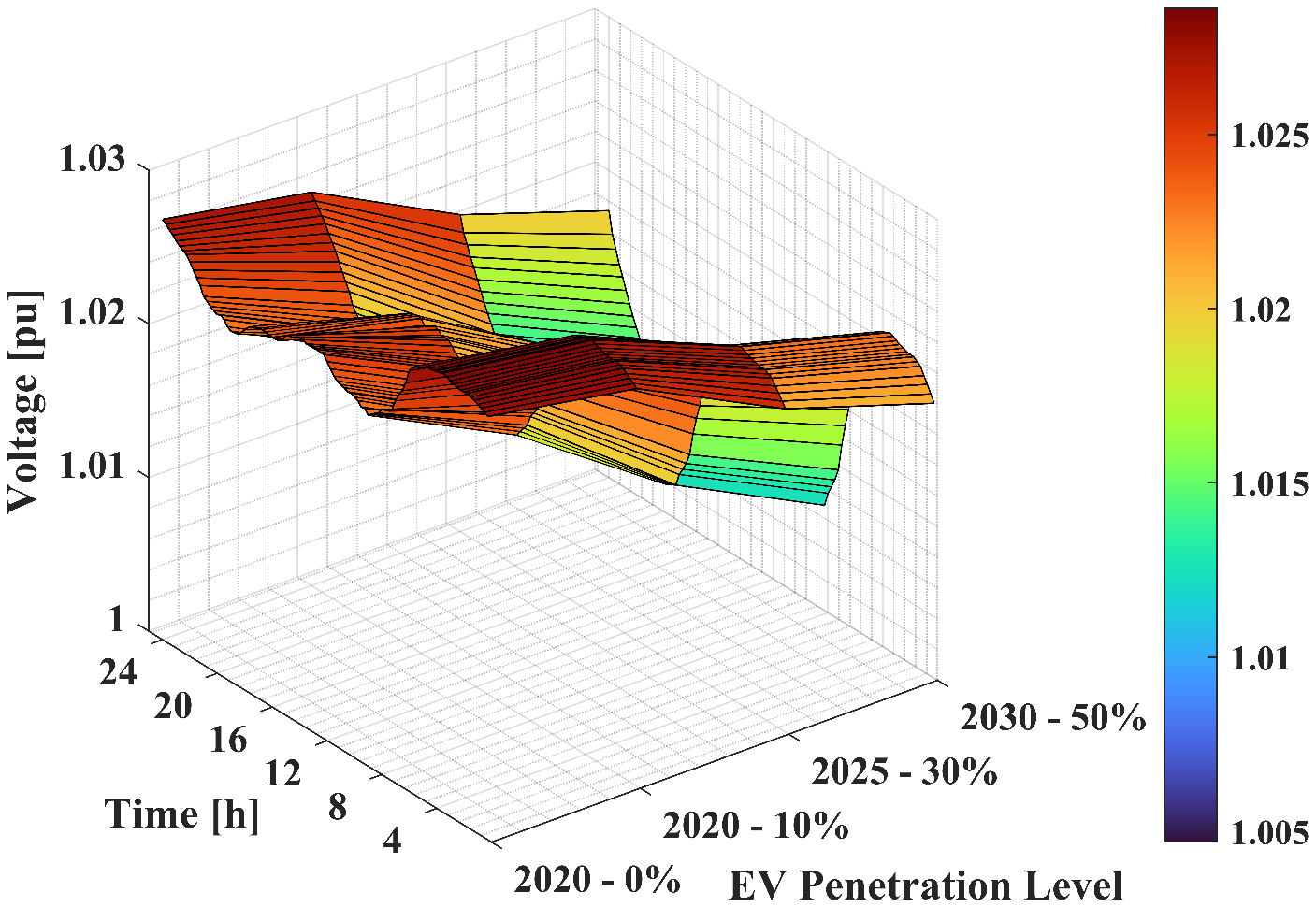}}
    &
    \subfloat[][Voltage profile (winter unoptimized).\label{fig:fig_15_2}]{\includegraphics[width=.46\columnwidth]{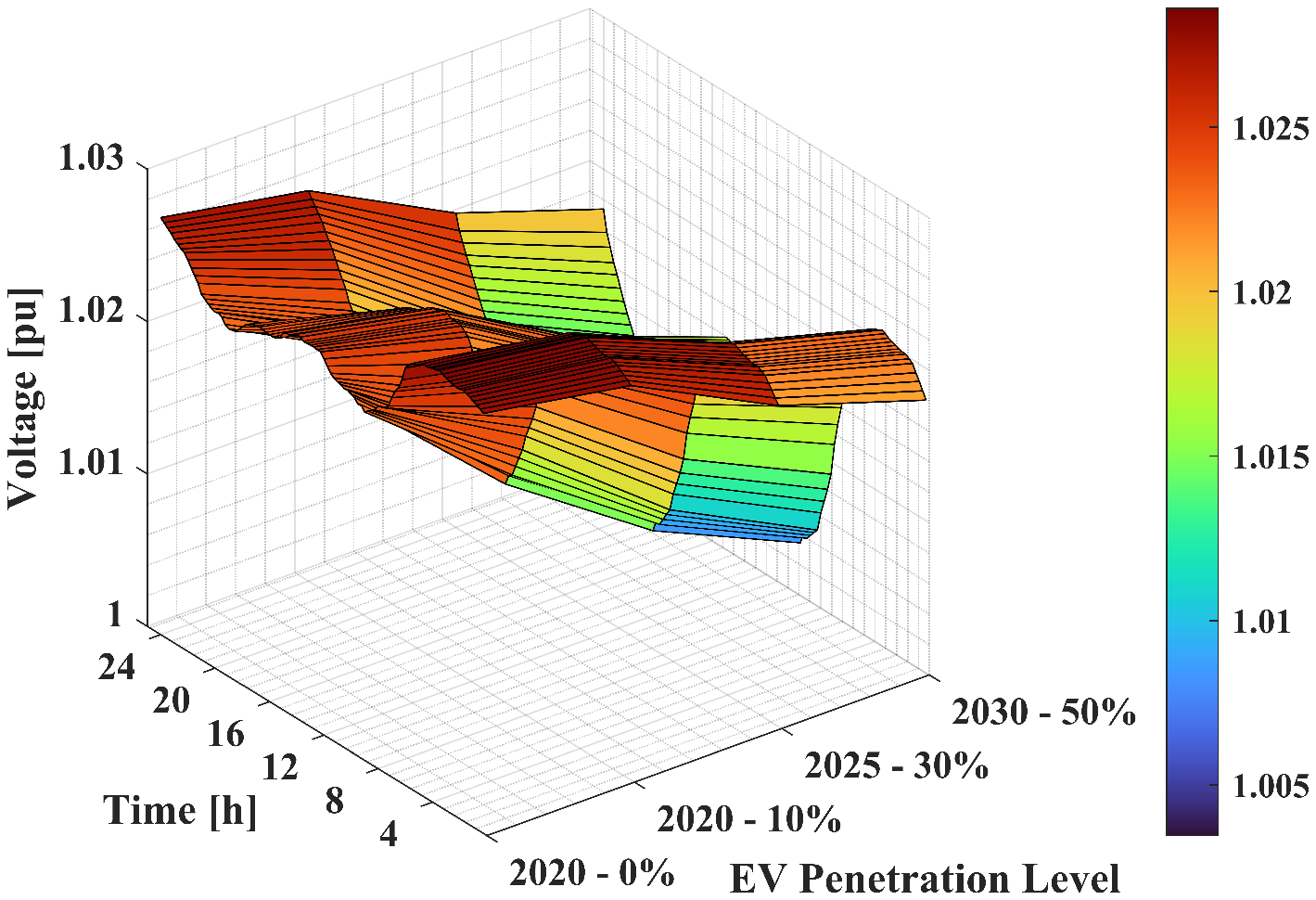}}\\
    \subfloat[][Trafo loading (summer optimized).\label{fig:fig_16_1}]{\includegraphics[width=.46\columnwidth]{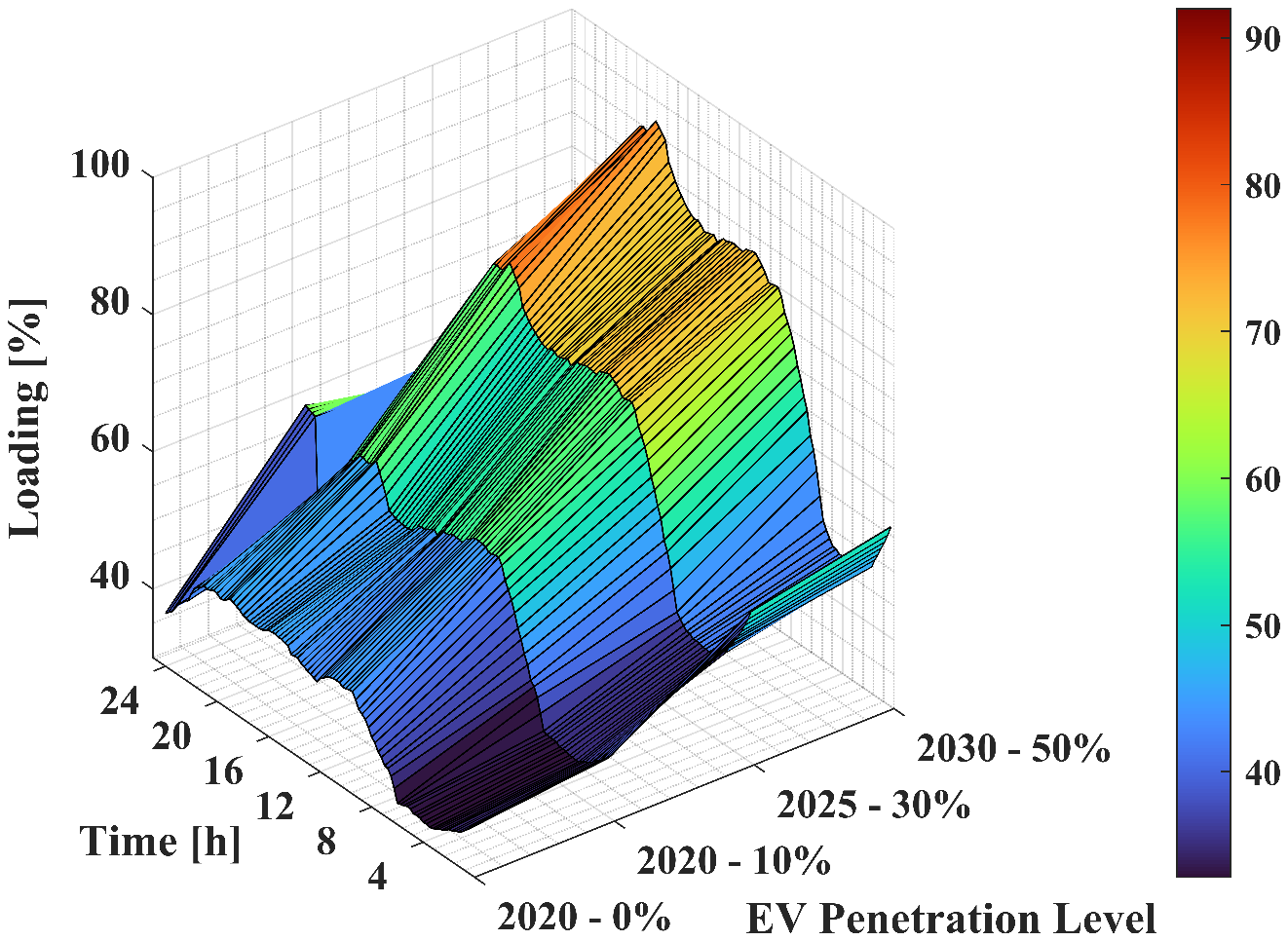}} &
    \subfloat[][Trafo loading (summer unoptimized).\label{fig:fig_16_2}]{\includegraphics[width=.46\columnwidth]{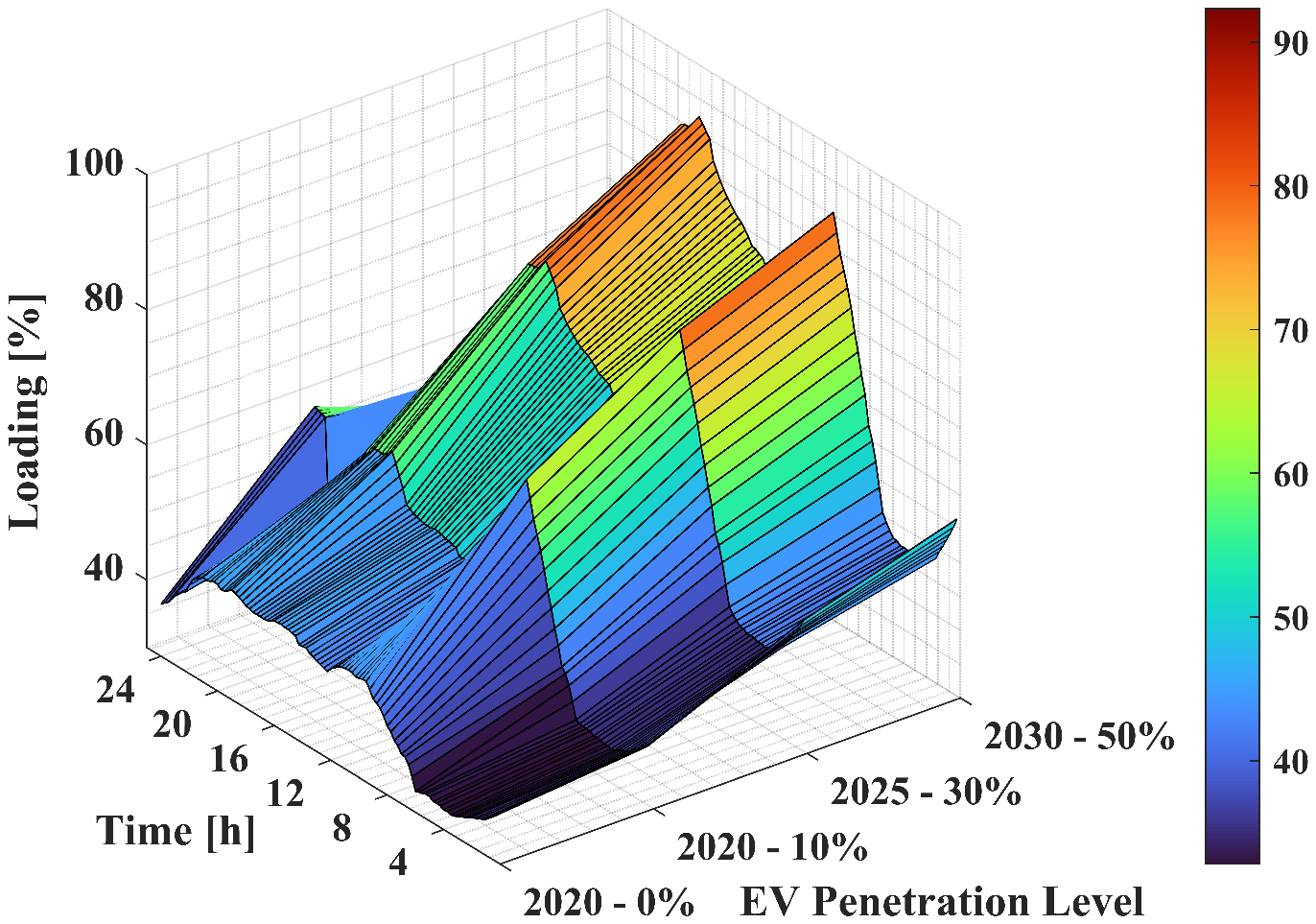}}
    &
    \subfloat[][Trafo loading (winter optimized).\label{fig:fig_17_1}]{\includegraphics[width=.46\columnwidth]{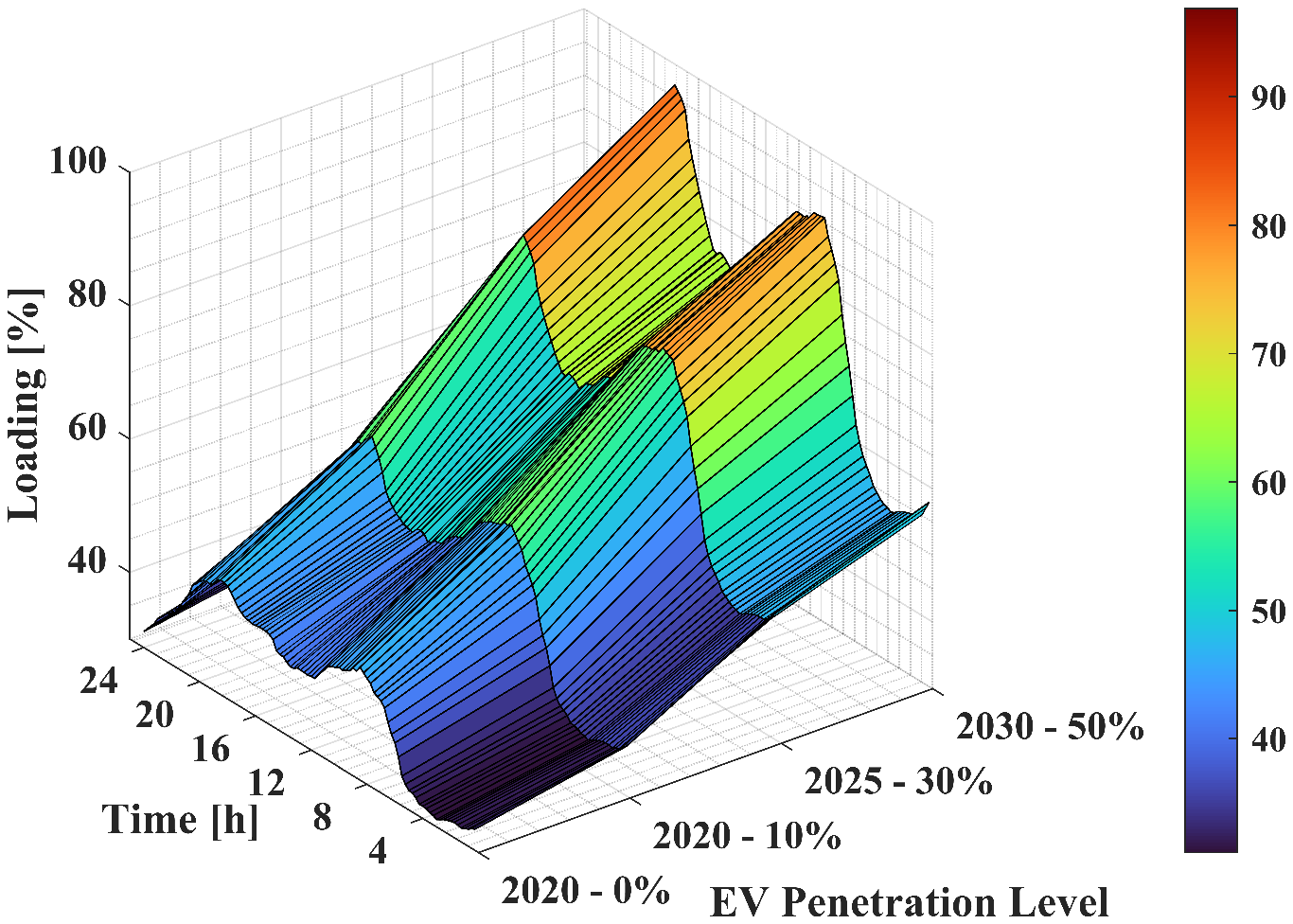}}
    &
    \subfloat[][Trafo loading (winter unoptimized).\label{fig:fig_17_2}]{\includegraphics[width=.46\columnwidth]{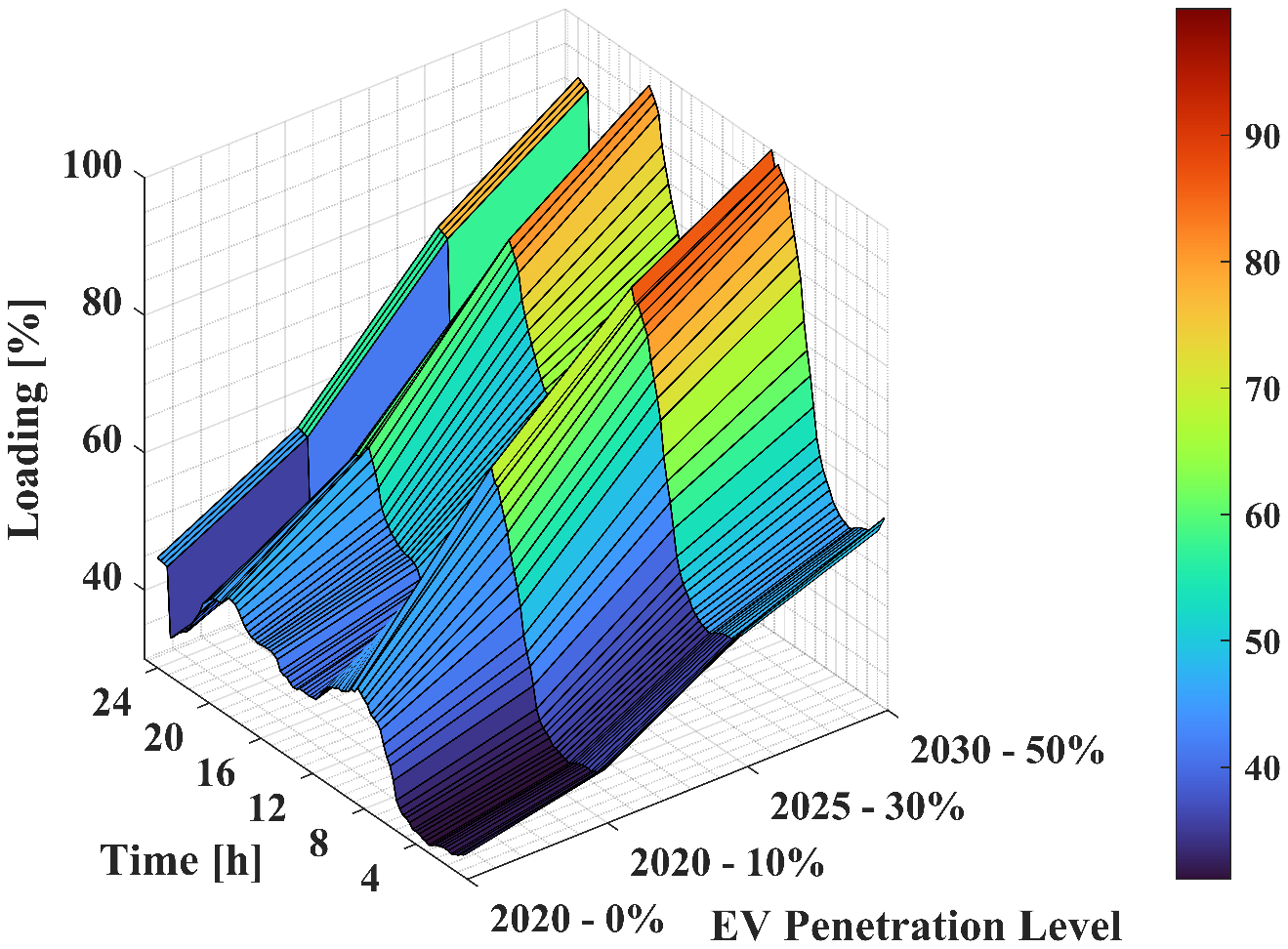}}\\
  \end{tabular}
  \caption{Plots of the impacts.}\label{fig:totale}
\end{figure*}

\subsection{Case I}

The study is conducted in winter and summer for the year 2020. The obtained result are shown in Figures~\subref{fig:fig_8_1}--\subref{fig:fig_9_2}. Numerical results are reported in Table~\ref{tab:tab_5}.

\begin{table}[ht]
  \centering
  \caption{Results obtained in Case I.}\label{tab:tab_5}
  \includegraphics[width=\columnwidth]{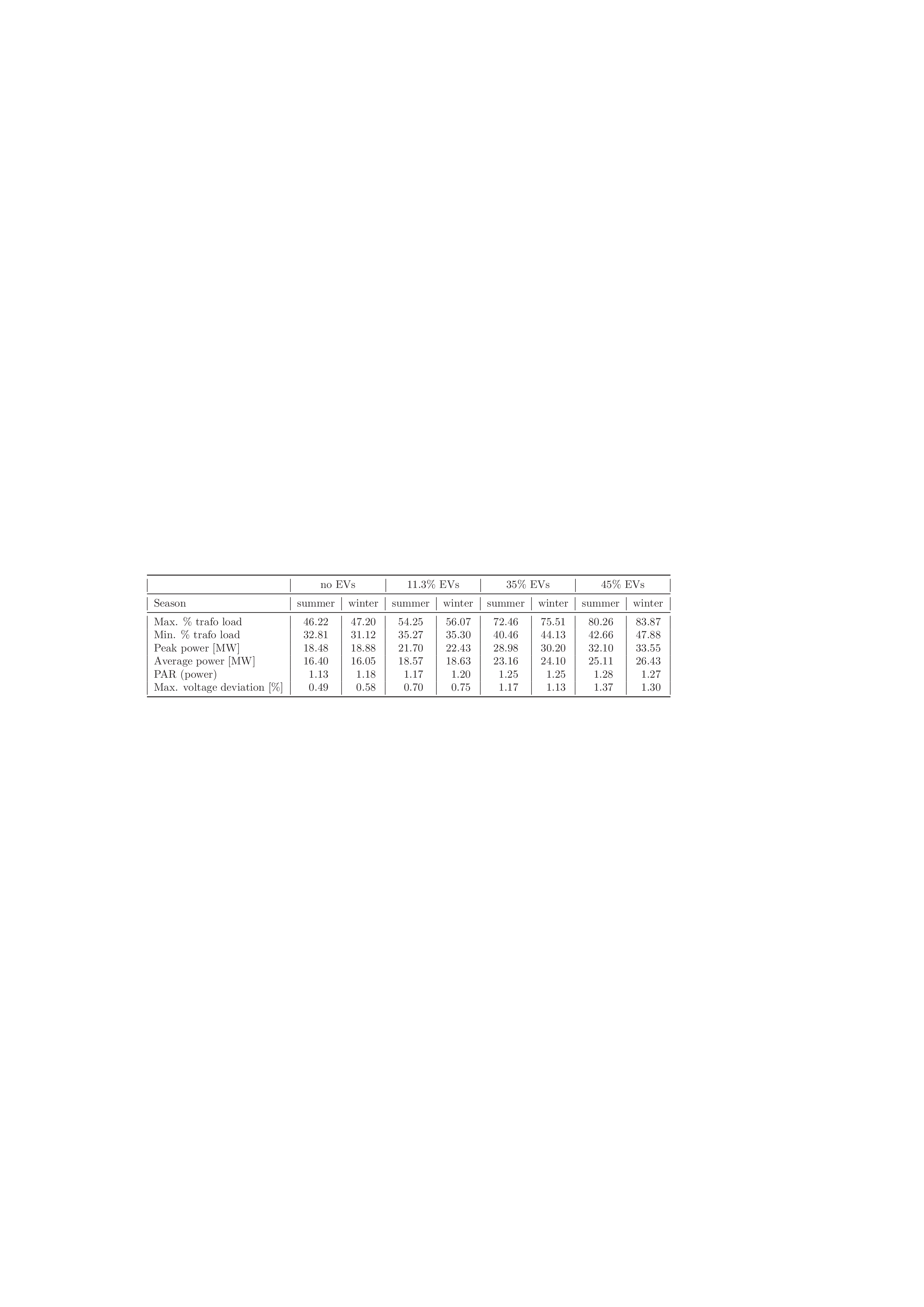}
\end{table}

Increasing in \acp{ev} load penetration caused an increase in voltage deviation; 45\% \ac{ev} penetration leads to about a 1\% increment in the maximum voltage deviation compared to no \ac{ev} penetration. However, this deviation can likely be limited to its cap, using voltage regulators in the network.

It illustrated that, during peak hours, the transformer is loaded up to 83.87\% in the worst case, in winter with 45\% \ac{ev} penetration. The grid is ready to accept up to about 40\% of \ac{ev} penetration, but above that the substitution of the main transformer is necessary.

\subsection{Case II}

The result obtained for Case II are shown in Figures~\subref{fig:fig_10_1}--\subref{fig:fig_11_2}. Numerical results are reported in Table~\ref{tab:tab_6}.

\begin{table}[h]
  \centering
  \caption{Results obtained in Case II.}\label{tab:tab_6}
  \includegraphics[width=\columnwidth]{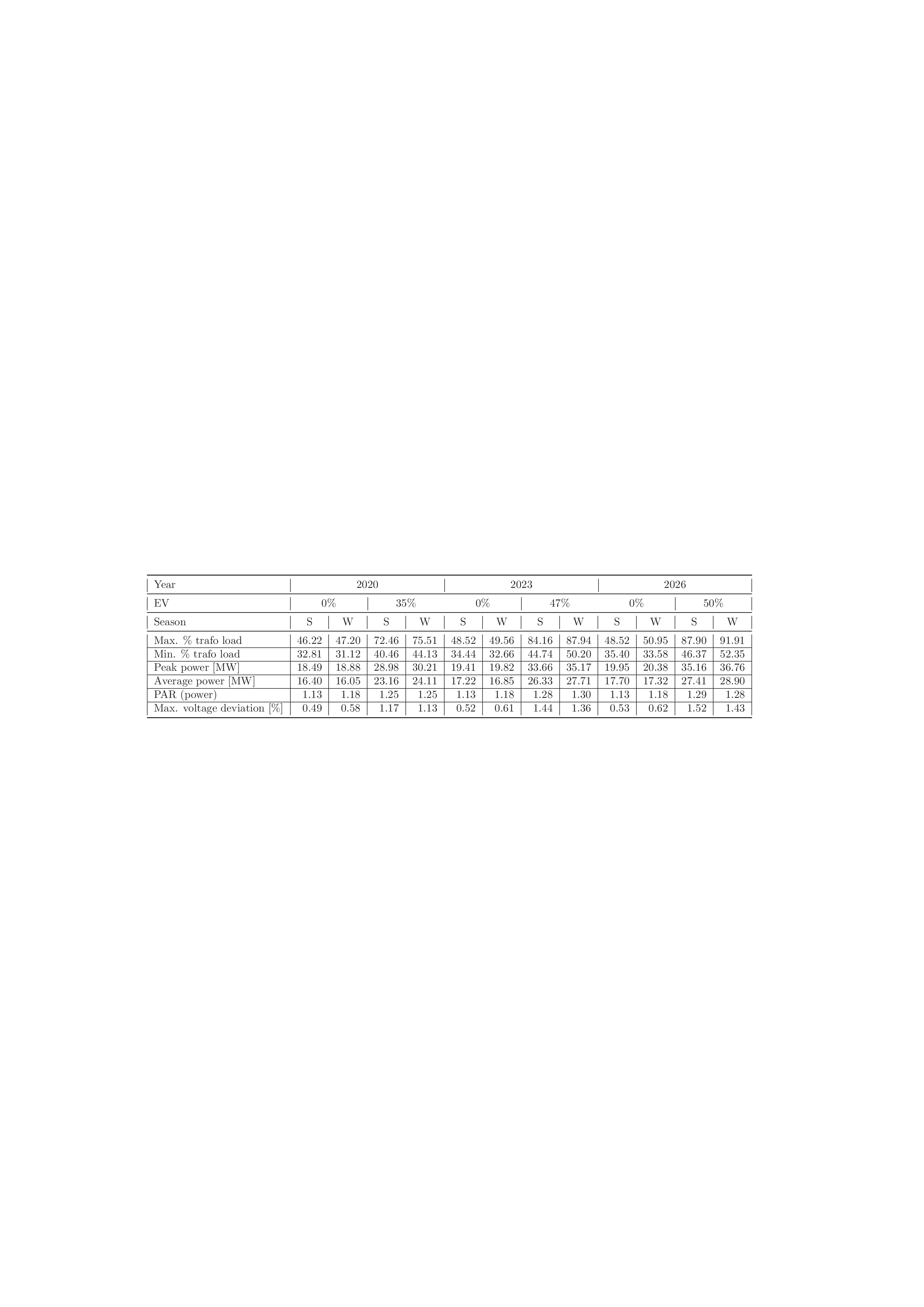}
\end{table}

After a careful assessment, the following findings can be reported:
The simulation results for this case lead to some practical considerations.
Both peak load and PAR are higher in 2023 than in 2020 but do not change significantly from 2023 to 2026. The reason for this is that the growth rate of \ac{ev} penetration level is higher within 2020--2023 than that of period 2023--2026 and the majority of EVs charge at the time of household peak load. Higher PAR will lead to high operation costs.

As expected, the increment of \ac{ev} charging stations causes an increment in the voltage deviation on the main busbar of the networks and also on each load nodes. Furthermore, the main transformer is loaded above 90\% and this will create aging on the equipment. The transformer should be reinforced with an additional cost. The cost of reinforcement can be minimized by optimizing the \ac{ev} loads.

Generally, it can be noted that winter is the worst case both in peak power and maximum voltage deviation.

\subsection{Case III}
The result obtained for Case III are shown in Figures~\subref{fig:fig_12_1}--\subref{fig:fig_13_2}. Numerical results are reported in Table~\ref{tab:tab_7}.

\begin{table}[h]
  \centering
  \caption{Results obtained in Case III.}\label{tab:tab_7}
  \includegraphics[width=\columnwidth]{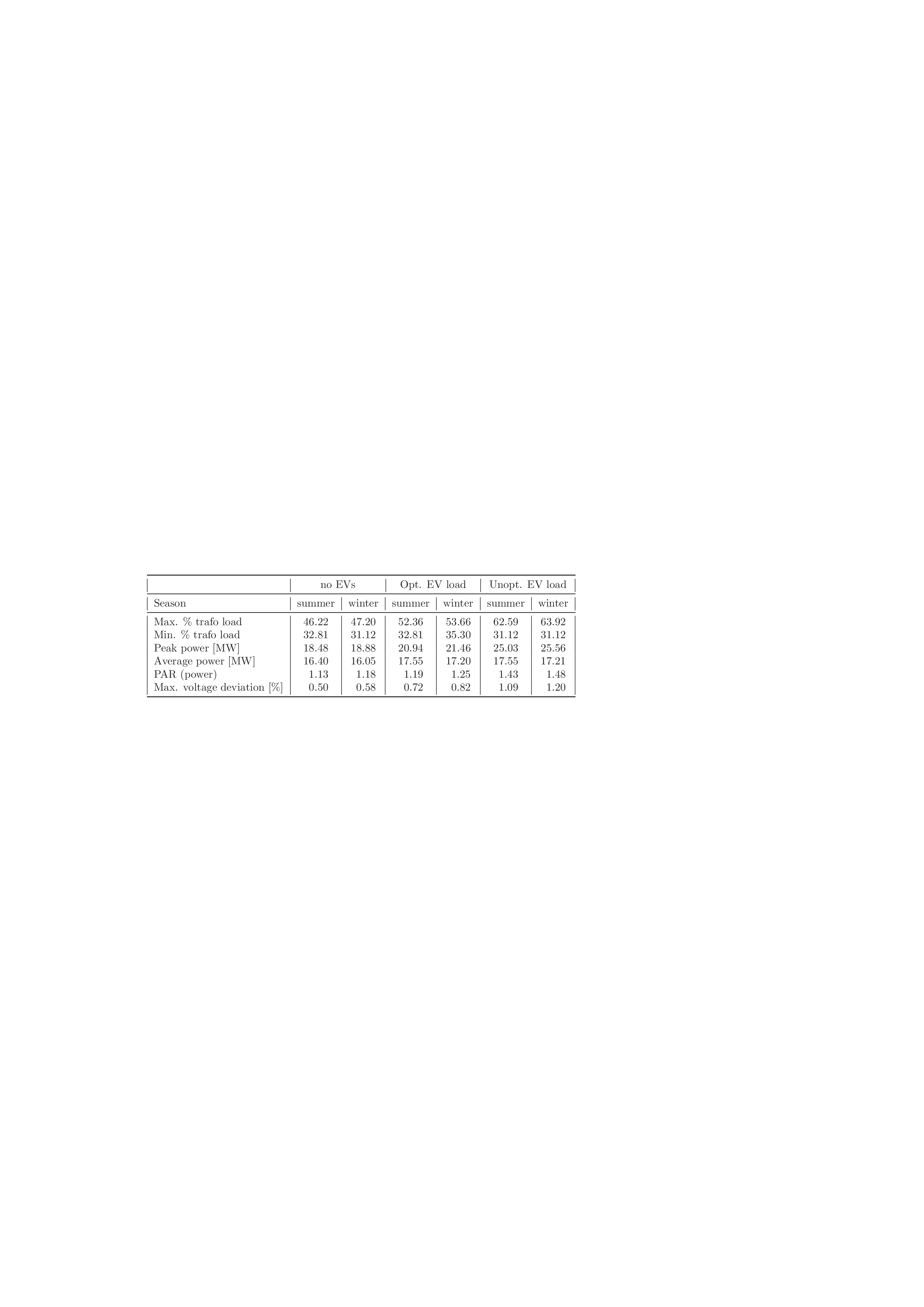}
\end{table}

The results show that a 16\% increase in the peak load when there is no load optimization. This will create considerable stress on the distribution grid and its elements even if it is for a short period of time. When the load optimization mechanism is applied the peak load increment, compared to the no-EV scenario, decreases from 16\% to only 6\% and the stress is reduced accordingly. By optimizing the load, the PAR is reduced from 1.48 to 1.25, in winter. This has a positive impact on reducing the operational stress of the power grid. The voltage profile is always been with in the acceptable limits according to EN50160.

\subsection{Case IV}
The result obtained for Case IV are shown in Figures~\subref{fig:fig_14_1}--\subref{fig:fig_17_2}. Numerical results are reported in Tables~\ref{tab:tab_8}--\ref{tab:tab_9}.

\begin{table}[ht]
  \centering
  \caption{Results obtained in Case IV - summer.}\label{tab:tab_8}
  \includegraphics[width=\columnwidth]{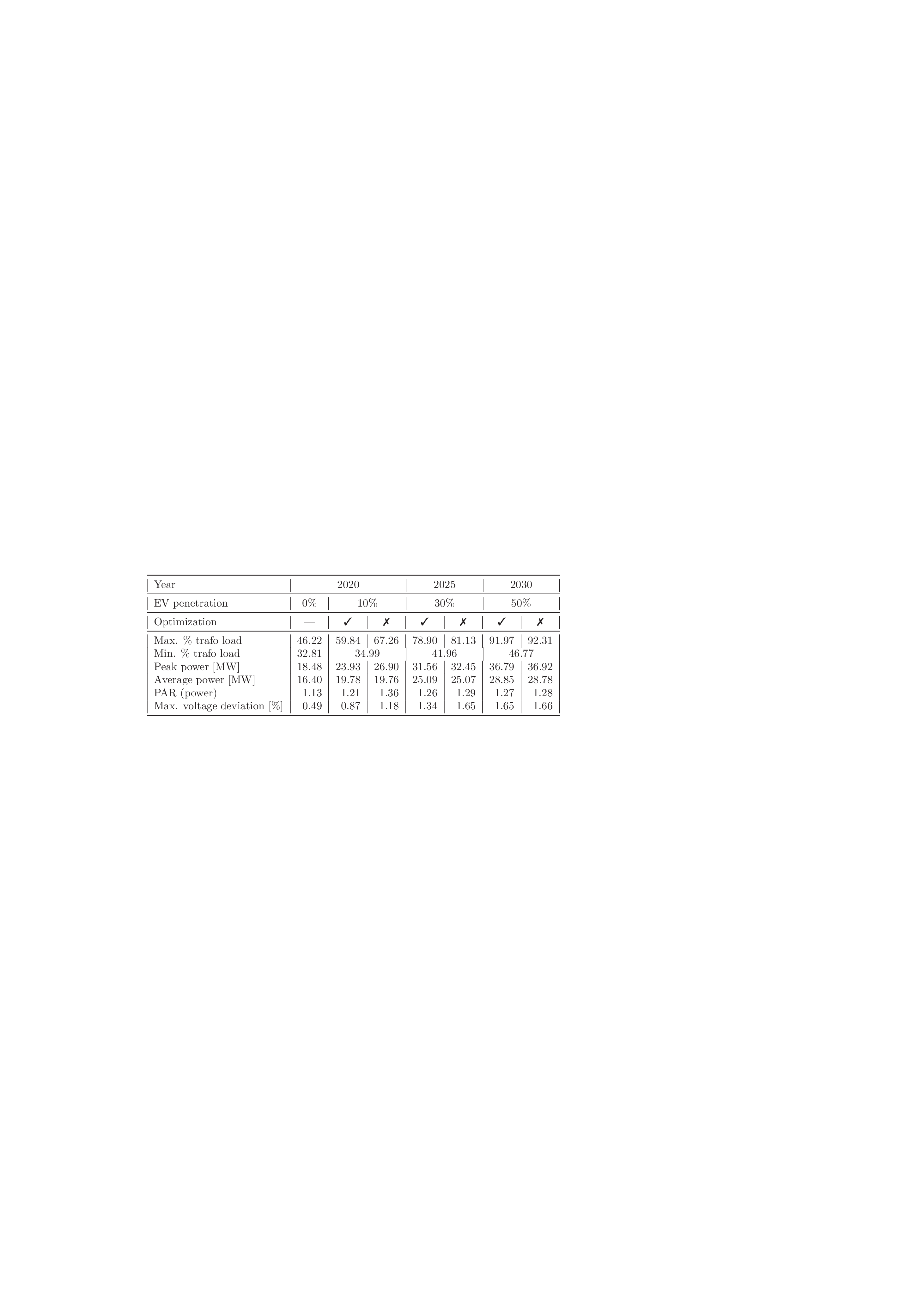}
\end{table}
\begin{table}[ht]
  \centering
  \caption{Results obtained in Case IV - winter.}\label{tab:tab_9}
  \includegraphics[width=\columnwidth]{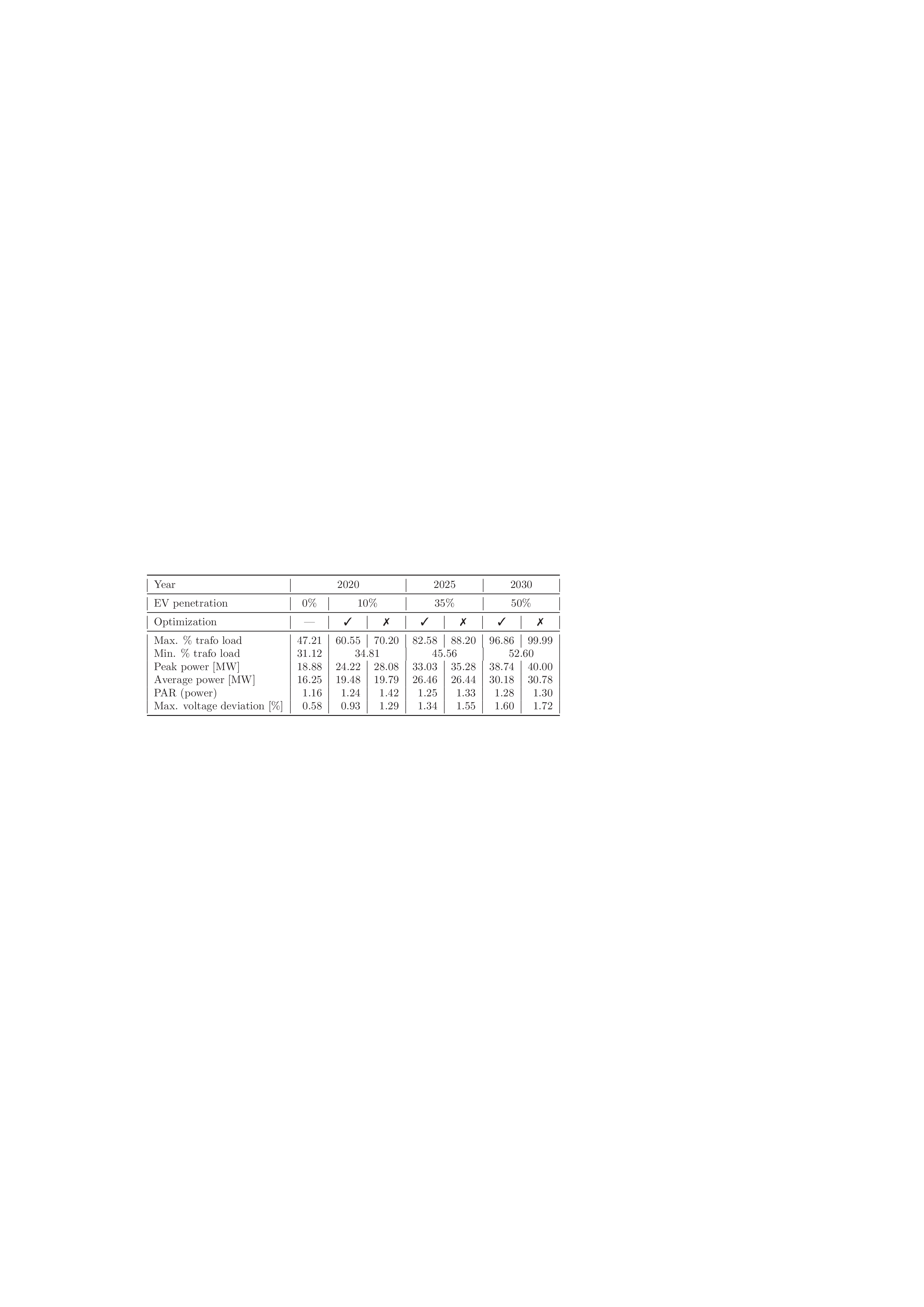}
\end{table}

The major problem found from this simulation is overloading of the main transformer. In 2030 in the winter season, the maximum loading of the transformer is 99.99\%. The transformer is already saturated, and it is not safe to work at this saturation level. On the unoptimized load scenarios, the loading graphs have very sharp edge on the peak hours. This shows higher difference between the peak power and lowest power which leads to higher PAR value.

\section{Conclusion}
In this paper, a detailed investigation of \ac{ev} charge load impacts was made on distribution systems in Italy and the mitigation mechanism is discussed. To realize this study, realistic data from a grid in Italy were used.

The study was performed in four different case studies with their own unique characteristics (i.e., \ac{ev} load penetration, years of study and type of \ac{ev} load). In order to have a more comprehensive view, the study was conducted for a summer and winter seasons.

The focus has been put primarily on the impact of \ac{ev} charging loads on the voltage profile, maximum voltage deviation, peak to average ratio (PAR) of active power, peak loads, and transformer loading. It was found that, as the \ac{ev} load penetration increases transformer loading and PAR increase. This results in unsuitability of the distribution network to accept high \ac{ev} load penetration. A slight voltage drop in the MV nodes was detected. Nevertheless, the voltage at MV nodes keeps lying within the acceptable limits. Hence, the main conclusion is that the network studied is ready, as it is, without further development, for a safe \ac{ev} load penetration up to 2026.

After implementation of the mitigation mechanism a considerable drop on both transformer loading and PAR was obtained, hence a better stability performance.\balance

The post COVID-19 situation could have a ``positive'' effect on fostering private transportation, as opposed to public transportation. More \ac{ev} charge penetration is expected in the future. This gives an alert that a special attention should be given to the issue studied in this paper.

\section*{Acknowledgment}
The research leading to this paper has been partially funded by the MIUR PRIN 2017K4JZEE\_006.


\end{document}